*US Nuclear Data Program*

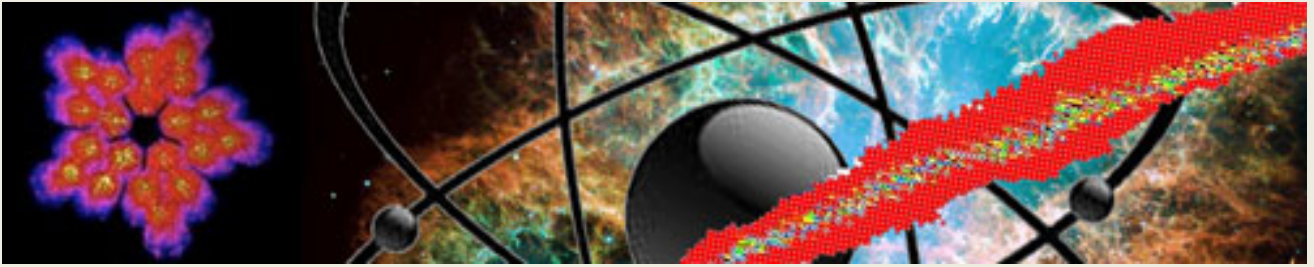

# White Paper on Nuclear Data Needs and Capabilities for Basic Science

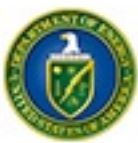

# Table of Contents

# List of Acronyms and Abbreviations

| | |
|---|---|
| **ARUNA** | Association for Research at University Nuclear Accelerators |
| **ATLAS** | Argonne Tandem Linac Accelerator System |
| **CASPAR** | Compact Accelerator System for Performing Astrophysical Research |
| **CCF** | Coupled Cyclotron Facility, NSCL |
| **DANCE** | Detector for Advanced Neutron Capture Experiments, LANL |
| **EDM** | Electric Dipole Moment |
| **ENDF** | Evaluated Nuclear Data File |
| **ENSDF** | Evaluated Nuclear Structure Data File |
| **EXFOR** | Experimental Nuclear Reaction Data |
| **FRIB** | Facility for Rare Isotope Beams, MSU |
| **G4LEND** | Geant4 Low Energy Nuclear Data software package for GEANT4 |
| **GEANT4** | Geometry and Tracking, Monte-Carlo particle transport code, CERN |
| **GRETINA** | Gamma-Ray Energy Tracking In-beam Nuclear Array, LBNL |
| **IAEA-NDS** | International Atomic Energy Agency – Nuclear Data Section |
| **JINA** | Joint Institute for Nuclear Astrophysics |
| **KADONIS** | Karlsruhe Astrophysical Database of Nucleosynthesis in Stars |
| **LANSCE** | Los Alamos Neutron Science Center |
| **LISE++** | Ligne d'Ions Super-Epluchés (Fragment separator simulation code) |
| **NACRE** | European Compilation of Reactions Rates for Astrophysics |
| **NEA** | Nuclear Energy Agency |
| **NDNCA** | Workshop on Nuclear Data Needs and Capabilities for Applications |
| **NNDC** | National Nuclear Data Center, BNL |
| **NSCL** | National Superconducting Cyclotron Laboratory, MSU |
| **NSDD** | Nuclear Structure and Decay Data Network, IAEA |
| **NSR** | Nuclear Science References |
| **PPAC** | Parallel Plate Avalanche Counter |
| **REACLIB** | Nuclear Reaction Rates Database, MSU |
| **SECAR** | Separator for Capture Reactions, MSU |
| **STARLIB** | Nuclear Physics Library for Astrophysical Research, UNC, Chapel Hill |
| **TAGS** | Total absorption $\gamma$-ray spectrometry |
| **TUNL** | Triangle University Nuclear Laboratory, Duke University |
| **USNDP** | United States Nuclear Data Program |
| **XUNDL** | Experimental Unevaluated Nuclear Data List |

# Rationale

The DOE Office of Science/Nuclear Physics -funded U.S. Nuclear Data Program (USNDP) comprises nuclear data experts from national laboratories and academia across the United States who collect, evaluate, and disseminate nuclear physics data for basic nuclear physics and applied nuclear technology research. The nuclear data infrastructure provided by the USNDP impacts governmental, educational, commercial, and medical organizations in United States, as illustrated in Figure 1, and it is a part of the U.S. commitment to various international nuclear data networks and collaborations.

Following on a recommendation by the USNDP Advisory Committee, a 1.5-day workshop entitled "***Nuclear Data Needs and Capabilities for Basic Science***" was organized on August 10-11, 2016 at the University of Notre Dame, in conjunction with the annual Low Energy Community Meeting. It followed the first workshop on *"Nuclear Data Needs and Capabilities for Applications"* (NDNCA) organized in May 2015 at the Lawrence Berkeley National Laboratory. The purpose of this second targeted workshop was to assemble and prioritize the needs of the nuclear physics research community for data sets, services and capabilities in areas including Nuclear Structure, Nuclear Reactions, Nuclear Astrophysics, Fundamental Interactions, Neutrino Physics and Nuclear Theory. More than 95 participants from 33 different institutions attended the workshop. The program is available at http://meetings.nscl.msu.edu/2016ND_workshop/html/program.html, together with copies of the presentations. An overview of nuclear data needs and capabilities identified at this meeting are summarized in the present document. Specific recommendations that can be used by the DOE Office of Science/Nuclear Physics office to guide future nuclear data activities are given along with the topical areas that directly benefit the broader nuclear physics community.

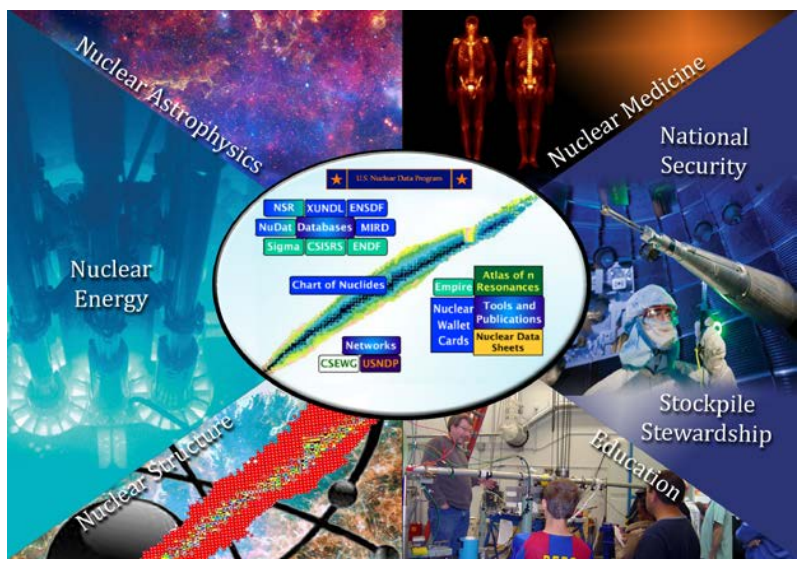

*Figure 1: The core USNDP nuclear physics databases and products, and their main areas of impact for science and technology.*

# Contributors

## Editors

J. Batchelder (UC Berkeley)
T. Kawano (LANL)
J. Kelley (NCSU & TUNL)
F.G. Kondev* (ANL)
E. McCutchan (BNL)
M. Smith (ORNL)
A. Sonzogni (BNL)
M. Thoennessen* (MSU)
I. Thompson (LLNL)
* Lead editors

## Overview speakers

B. Balantekin (UWM)
A. Brown (MSU)
R. Casten (Yale U)
A. Champagne (UNC)
H. Crawford (LBNL)
L. McCutchan (BNL)
G. Savard (ANL/U Chicago)
B. Sherrill (MSU)
L. Sobotka (WUSL)

## Contributed speakers

M. Almond (ORNL)
L. Bernstein (LBNL)
M.P. Carpenter (ANL)
J. Clark (ANL)
J. Greene (ANL)
D.J. Hartley (USNA)
B. Kay (ANL)
T. Koi (SLAC)
F.G. Kondev (ANL)
H-Y. Lee (LANL)
S. Liddick (MSU)
E. Olsen (MSU)
G. Perdikakis (CMU)
M. Redshaw (MSU)
A. Rogers (UML)
K. Rykaczewski (ORNL)
H. Schatz (MSU)
M. Smith (ORNL)
O. Tarasov (MSU)
C-Y. Wu (LLNL)

## Acknowledgments

The work performed by staff from several laboratories was sponsored by the US DOE Office of Science, Office of Nuclear Physics under Contracts DE-AC02-06CH11357 (ANL), DE-AC02-98CH10886 (BNL), DE-AC02-05CB11231 (LBNL), DE-SC0016948 (MSU), DE-AC05-76OR00022 (ORNL), and DE-FG02-97ER41033 (TUNL). The work by staff from Lawrence Livermore National Laboratory and Los Alamos National Laboratory was sponsored by the US DOE NNSA under Contract DE-AC52-07NA27344 and DE-AC52-06NA25396 respectively. The authors wish to thank Dr. P. Dimitriou (IAEA) for her contributions to the document.

# Executive Summary & Conclusions

Reliable nuclear structure and reaction data represent the fundamental building blocks of nuclear physics and astrophysics research, and are also of importance in many applications. There is a continuous demand for high-quality updates of the main nuclear physics databases via the prompt compilation and evaluation of the latest experimental and theoretical results. The nuclear physics research community benefits greatly from comprehensive, systematic and up-to-date reviews of the experimentally determined nuclear properties and observables, as well as from the ability to rapidly access these data in user-friendly forms. Such credible databases also act as a bridge between science, technology, and society by making the results of basic nuclear physics research available to a broad audience of users, and hence expand the societal utilization of nuclear science. Compilation and evaluation of nuclear data has deep roots in the history of nuclear science research, as outlined in Appendix 1. They have an enormous impact on many areas of science and applications, as illustrated in Figure 2 for the Evaluated Nuclear Structure Data File (ENSDF) database.

The first workshop on *"Nuclear Data Needs and Capabilities for Applications"* (NDNCA) presented a comprehensive overview of the nuclear data needs for applications and proposed a plan to coordinate a framework in line with the existing Nuclear Data High Priority List of the Nuclear Energy Agency [1] to assess such data needs. In response, an ad-hoc Nuclear Data Working group was formed which subsequently proposed a plan to address the most important data needs, as specified in the workshop white paper [2]. The NDNCA activities focussed on the data needs of the applications communities which are primarily concentrated on a need for improving the databases by obtaining new data that clarifies regions where incomplete or unreliable data are currently present; such improvements would positively impact results of calculations and simulations that rely on nuclear data input. Capabilities in the context of the NDNCA workshop refer to experimental facilities, as nicely compiled in an appendix of the applications white paper [2].

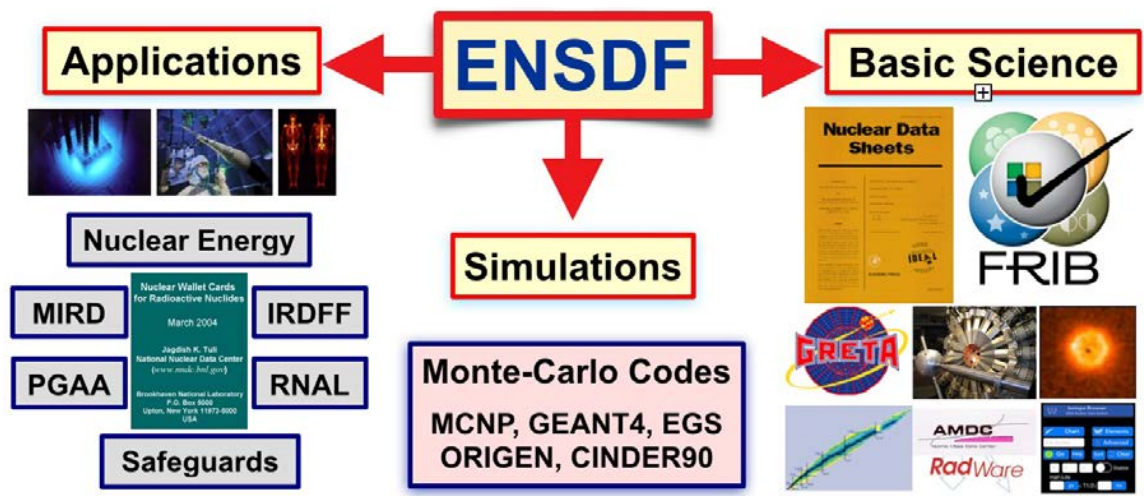

*Figure 2: Impact of the evaluated Nuclear Structure Data File (ENSDF) on basic nuclear science and broader applications*

The present workshop concentrated on the needs of the basic nuclear science community for data and capabilities. The main role of this community is to generate and use data in order to understand the basic nuclear forces and interactions that are responsible for the existence and the properties of all nuclides and, as a consequence, to gain knowledge about the origins, evolution and structure of the universe. Thus, the experiments designed to measure a wealth of nuclear properties towards these fundamental scientific goals are typically performed from within this community.

For the basic nuclear science community *"data needs"*, in the context of the USNDP mission, refers to the need for comprehensive and up-to-date evaluated nuclear science databases, which serve as the foundation of theoretical models and for identifying and planning future experimental and theoretical studies. For this community, *"data capabilities"* are essentially the tools available to interrogate, interpret and analyse the evaluated nuclear data from these databases.

The USNDP plays a seminal role within the basic nuclear physics community and, therefore, close connections and communications between the researchers and the data evaluators are essential. We advocate that nuclear data evaluation should be the final step for any experiment, since only after the results are compiled and evaluated it can be ensured that the knowledge gained advances science. In this context, the ultimate goal of every experiment should be the inclusion of the acquired data into the evaluated nuclear physics databases. A failure to do so will represent a waste of resources that were expended during the planning, preparation, execution and analysis stages of the experiment.

Evaluated nuclear data are critical for the advancement of the nuclear physics field. Although individual experiments can lead directly to important discoveries, it is also recognized that major advances in nuclear science can be driven by systematic studies of compiled results from many different experiments. This can lead to the development of new theoretical interpretations and models, which in turn inspire new experiments in order to test the predictive power of the models. In the near future, the modern nuclear physics facilities around the world, and the Facility for Rare Isotope Beams (FRIB) in particular, will provide unprecedented access to a vast new terrain of nuclei. The scientific breakthroughs and major advances in our understanding of nuclei and their role in the cosmos that will be enabled by these facilities will only be possible through more resources, new tools and methodologies, and increased community involvement to handle the expected deluge of new data.

# Conclusions

The workshop participants expressed overwhelming and unequivocal support for the existence, maintenance and future development of a nuclear structure database (ENSDF). Specifically:

- *The evaluated data should be reliable, comprehensive and up-to-date. To achieve this goal there should be continuous funding support for the existing data evaluators and an expansion of the pool of skilled nuclear structure data evaluators is imperative for succession planning.*

    Several speakers expressed concerns about the present status of ENSDF, while many others inherently assumed that ENSDF is up-to-date and did not realize the amount of effort needed to maintain the quality and currency of the database. The needs that they discussed are only valid and useful with the assumption that the underlying database is a) reliable and credible – data must be correctly evaluated; b) comprehensive – should include all measured quantities and their uncertainties; c) up-to-date – results from all measurements should be promptly incorporated; and d) accessible – easy and rapid availability in user-required formats. Those are the foundation principles that should guide the future development of the USNDP nuclear structure and reaction databases.

    The current facilities, including ATLAS, the CCF at the NSCL, and the ARUNA laboratories will continue to generate new data that need to be compiled and evaluated. Data generation will significantly increase even more when FRIB becomes operational. Importantly, these facilities will not only produce new data, but they will also use the available evaluated data for the planning and preparation of new experiments. Therefore, the timely compilation and evaluation of nuclear structure and reaction data are essential and of vital importance.

- *Capabilities for the compilation and evaluation of new and more complex data types should be developed.*

    In the FRIB era many new types of data, including data with increased complexity, will be generated, which will require upgrades of current database formats and policies. For example, $\gamma$-ray strength function data and results from calorimetric $\gamma$-ray spectroscopy studies are currently not uniformly incorporated in ENSDF. Nevertheless, they provide important nuclear structure information about the properties of excited states.

- *Connections to nuclear astrophysics research needs to be strengthened and expanded*

    The interdisciplinary field of nuclear astrophysics has extensive data needs in both reaction and structure physics. It also requires specialized data processing steps in order to enable this data to be used as critical input for simulations of cosmic systems. Current efforts in this area are subcritical, however. To maximize the scientific return on recent facility investments for measurements in this area, USNDP activities should be expanded to include efforts in evaluations, databases, and tools specifically targeted for nuclear astrophysics.

- *Connections to theoretical databases should be established.*

  Progress in all areas of nuclear physics requires the critical comparison of theoretical predictions to experimental data. Theoretical models have, however, greatly expanded their predictive power in scope and complexity, and the number of groups producing such sophisticated data sets has also expanded. The USNDP should explore the establishment of databases and tools necessary to facilitate the comparison of large theoretical nuclear datasets with evaluated nuclear data.

- *Accessibility to the databases should be improved.*

  The generation of comprehensive and up-to-date databases by itself is not sufficient to fully exploit the potential of scientific discoveries. Appropriate interfaces that allow the users to easily interrogate the evaluated data are required. In close collaboration with the nuclear physics research community, USNDP should develop innovative software tools for display, extraction and manipulation of the evaluated data. In addition, establishing a version-controlled publication of ENSDF would allow unambiguous citations of reproducible quantities.

- *Compilation of new data should be ensured.*

  From the data needs expressed at the workshop, it is apparent that the demand on USNDP is considerable. The basic nuclear science community should also take on a greater responsibility in the compilation of data they produce, which would allow the data scientists to concentrate their efforts on the evaluation process. Thus, it is imperative that the experimenters publish all data in sufficient detail and in a readable format so they can be easily incorporated into the databases. Data-related pre-review of journal manuscripts is encouraged and should be pursued. Finally, there is a large amount of historical data which have not been evaluated, because it is not available in a directly readable format. In many cases, these data are relevant to current research activities and it would be beneficial to recover them. This would be much more cost effective than to repeat the experiments. For some data, digitizing the old results is the only option, because the experimental capabilities for taking these data do not exist anymore.

[1] NEA Nuclear Data High Priority Request List, https://www.oecd-nea.org/dbdata/hprl/.
[2] Nuclear Data Needs and Capabilities for Applications, May 27-29, 2015, Lawrence Berkeley National Laboratory, Berkeley CA, arXiv:1511.07772.

# Participants

1. M. Almond — Oak Ridge National Laboratory
2. A. Aprahamian — University of Notre Dame
3. B. Balantekin — University of Wisconsin Madison
4. B. Baramsai — Los Alamos National Laboratory
5. D. Bardayan — University of Notre Dame
6. T. Barnes — DOE-NP
7. J. Batchelder — University of California Berkeley
8. M. Beard — University of Notre Dame
9. I. Bentley — Saint Mary's College
10. L. Bernstein — Lawrence Berkeley National Laboratory
11. T. Bredeweg — Los Alamos National Laboratory
12. N. Brewer — ORNL, University of Tennessee
13. M. Brodeur — University of Notre Dame
14. A. Brown — Michigan State University
15. C. Brune — Ohio University
16. D. Burdette — University of Notre Dame
17. M. Caprio — University of Notre Dame
18. M. Carpenter — Argonne National Laboratory
19. R. Casten — Yale University
20. A. Champagne — University of North Carolina at Chapel Hill & TUNL
21. J. Chen — Michigan State University
22. D. Chrisman — Michigan State University
23. A. Clark — University of Notre Dame
24. J. Clark — Argonne National Laboratory
25. V. Constatinou — University of Notre Dame
26. N. Cooper — University of Richmond
27. A. Couture — Los Alamos National Laboratory
28. H. Crawford — Lawrence Berkeley National Laboratory
29. J.J. Das — Michigan State University
30. M. Devlin — Los Alamos National Laboratory
31. P. DeYoung — Hope College
32. M. Famiano — Western Michigan University
33. N. Frank — Augustana College
34. J. Freeman — Hampton University

| | | |
|---|---|---|
| 35. | N. Gamage | Central Michigan University |
| 36. | D. Garand | Michigan State University |
| 37. | S. Garrett | Lawrence Berkeley National Laboratory |
| 38. | J. Greene | Argonne National Laboratory |
| 39. | U. Greife | Colorado School of Mines |
| 40. | P. Gueye | Hampton University |
| 41. | D. Hartley | US Naval Academy |
| 42. | S. Henderson | University of Notre Dame |
| 43. | C. Hoffman | Argonne National Laboratory |
| 44. | G. Imbriani | University of Notre Dame |
| 45. | T. Kawano | Los Alamos National Laboratory |
| 46. | B. Kay | Argonne National Laboratory |
| 47. | J. Kelley | North Carolina State University & TUNL |
| 48. | J. Kelly | University of Notre Dame |
| 49. | T. Koi | SLAC |
| 50. | F. Kondev | Argonne National Laboratory |
| 51. | S. Lesher | University of Wisconsin at LaCrosse |
| 52. | X. Li | University of Notre Dame |
| 53. | S. Liddick | Michigan State University |
| 54. | H. Liu | Michigan State University |
| 55. | S. Marley | Louisiana State University |
| 56. | T. Massey | Ohio University |
| 57. | A. McCoy | University of Notre Dame |
| 58. | E. McCutchan | Brookhaven National Laboratory |
| 59. | L. Morales | University of Notre Dame |
| 60. | S. Mosby | Los Alamos National Laboratory |
| 61. | A. Nelson | University of Notre Dame |
| 62. | E. Olsen | Michigan State University |
| 63. | G. Perdikakis | Central Michigan University |
| 64. | W. Porter | University of Notre Dame |
| 65. | C. Pruitt | Washington University at St. Louis |
| 66. | T. Redpath | Michigan State University |
| 67. | M. Redshaw | Central Michigan University |
| 68. | C. Reidhead | Brigham Young University |
| 69. | M. Riley | Florida State University |
| 70. | A. Rogers | University of Massachusetts Lowell |
| 71. | W. Rogers | Indiana Wesleyan University |
| 72. | K. Rykaczewski | Oak Ridge National Laboratory |
| 73. | G. Savard | Argonne National Laboratory |
| 74. | H. Schatz | Michigan State University |
| 75. | B. Schultz | University of Notre Dame |
| 76. | J. Sethi | University of Maryland & Argonne National Laboratory |
| 77. | B. Sherrill | Michigan State University |
| 78. | A. Simon | University of Notre Dame |
| 79. | J. Smith | University of Connecticut |
| 80. | M. Smith | Oak Ridge National Laboratory |
| 81. | L. Sobotka | Washington University Saint Louis |
| 82. | A. Sonzogni | Brookhaven National Laboratory |
| 83. | S. Strauss | University of Notre Dame |
| 84. | R. Talwar | Argonne National Laboratory |
| 85. | W. Tan | University of Notre Dame |
| 86. | O. Tarasov | Michigan State University |
| 87. | M. Thoennessen | Michigan State University |
| 88. | I. Thompson | Lawrence Livermore National Laboratory |
| 89. | B. Vande Kolk | University of Notre Dame |
| 90. | A. Villari | Michigan State University |
| 91. | D. Votaw | Michigan State University |
| 92. | M. Wiescher | University of Notre Dame |
| 93. | J. Winkelbauer | Los Alamos National Laboratory |
| 94. | C. Wu | Lawrence Livermore National Laboratory |
| 95. | S. Yates | University of Kentucky |
| 96. | S. Zhu | Argonne National Laboratory |

# Program

## August 10, 2016
## Session 1 - Chair: M. Thoennessen (MSU)

| | |
|---|---|
| 8:40-8:45 am | Welcome & Opening<br>*A. Aprahamian, UND* |
| 8:45-9:00 am | Comments from DOE-NP<br>*T. Barnes, DOE/SC/NP* |
| 9:00-9:35 am | Nuclear Data – overview<br>*L. McCutchan, BNL* |
| 9:35-10:00 am | Nuclear Data Capabilities and Needs: Atomic Masses & Other Horizontal Evaluations<br>*F. Kondev, ANL* |
| 10:00-10:30 am | Coffee Break |

## Session 2 - Chair: M. Smith (ORNL)

| | |
|---|---|
| 10:30-11:00 am | Fundamental Interaction - overview<br>*G. Savard, ANL/UC* |
| 11:00-11:15 am | How to best evaluate and disseminate data of states above (continuum) and below (quasicontinuum) particle separation energy<br>*L. Bernstein, LBNL/UCB* |
| 11:15-11:30 am | Measurements of total prompt $\gamma$-ray energy distributions in neutron-induced fissions using DANCE<br>*Ching-Yen Wu, LLNL* |
| 11:30-11:45 am | On-going nuclear data activities at LANSCE<br>*Hye Young Lee, LANL* |
| 11:45-12:00 pm | Experimental nuclear-physics program at UMass Lowell<br>*A. Rogers, UML* |
| 12:00-12:15 pm | Nuclear Data Measurements in the Physics Division at ANL<br>*J. Greene, ANL* |
| 12:15-1:30 pm | Lunch |

## Session 3 - Chair: L. Bernstein (LBNL)

| | |
|---|---|
| 1:30-2:00 pm | Nuclear Astrophysics<br>*A. Champagne, UNC* |
| 2:00-2:15 pm | Current efforts, deficiencies, and future opportunities related to astrophysical reaction rate evaluation<br>*H. Schatz, MSU* |
| 2:15-2:30 pm | Software Systems for the U.S. Nuclear Data Program<br>*M. Smith, ORNL* |
| 2:30-2:45 pm | Statistical model reaction rates for the synthesis of heavy elements in stars<br>*G. Perdikakis, CMU* |
| 2:45-3:00 pm | How accurate mass data is crucial for models of the astrophysical r-process<br>*J. Clark, ANL* |
| 3:00-3:15 pm | Nuclear energy level and mass data for studies of rare and weak $\beta$ decays<br>*M. Redshaw, CMU* |
| 3:15-3:30 pm | Compilation and assessment of transfer-reaction data<br>*B. Kay, ANL* |
| 3:30-4:00 pm | Coffee Break |

## Session 4 - Chair: I. Thompson (LLNL)

| | |
|---|---|
| 4:00-4:30 pm | Nuclear Theory - overview<br>*B.A. Brown, MSU* |
| 4:30-4:45 pm | Massexplorer website: calculations of ground state nuclear properties<br>*E. Olsen, MSU* |
| 4:45-5:15 pm | FRIB - overview<br>*H. Crawford, LBNL* |
| 5:15-5:30 pm | Nuclear Data needs for LISE++<br>*O. Tarasov, MSU* |
| 5:30-5:45 pm | How Public Data Libraries Are Used in Geant4<br>*T. Koi, SLAC* |
| 5:45-6:15 pm | Open Discussions from the floor |

# August 11, 2016

## Session 1 - Chair: J. Kelley (NCS/TUNL)

| | |
|---|---|
| 9:00-9:30 am | Neutrinos – overview<br>*B. Balantekin, UWM* |
| 9:30-9:45 am | Data important for the nuclear fuel cycle and anti-neutrino physics obtained with Modular Total Absorption Spectrometer MTAS at ORNL<br>*K. Rykaczewski, ORNL* |
| 10:00-10:15 am | Nuclear Data Needs for Experimental Studies of Excited States in Nuclei by $\gamma$-Ray Spectroscopy<br>*M.P. Carpenter, ANL* |
| 10:15-10:45 am | Nuclear Reactions – overview<br>*L. Sobotka, WUSL* |
| 10:45-11:15 am | Coffee Break/Workshop Photo |

## Session 2 - Chair: M. Riley (FSU)

| | |
|---|---|
| 11:15-11:45 am | Nuclear Structure – overview<br>*R. Casten, Yale U* |
| 11:45-12:00 pm | The importance of the Nuclear Data program to research<br>*D.J. Hartley, USNA* |
| 12:00-12:15 pm | The efficiencies of evaluated nuclear data<br>*S. Liddick, MSU* |
| 12:15-12:30 pm | Data Evaluation Needs for Experimental and Theoretical Investigations of Nuclear Structure<br>*M. Almond, ORNL* |
| 12:30-1:00 pm | Summary and Conclusions<br>*B. Sherrill, MSU* |
| 1:00-1:10 pm | Closing<br>*M. Thoennessen, MSU*<br>*F. Kondev, ANL* |

# Workshop photograph

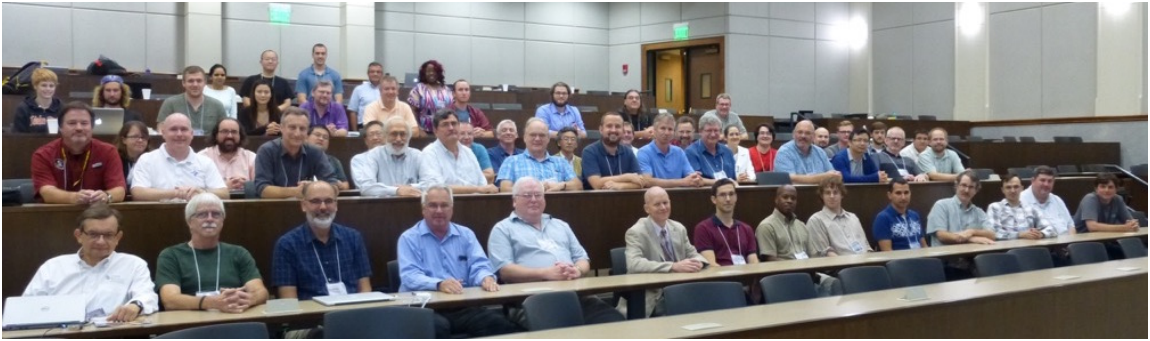

# 1. Nuclear Structure

The community showed overwhelming support for the Nuclear Science Reference (NSR), Evaluated Nuclear Structure Data File (ENSDF), and eXperimental Unevaluated Nuclear Data List (XUNDL) databases. The structure databases and services of the USNDP were referred to as "an incredibly valuable resource" and it was clear that "ENSDF and XUNDL must continue" in order for the nuclear structure community to operate. A schematic diagram of the lifecycle of a nuclear structure experiment is given in Figure 3. Certainly, one main outcome of an experiment should be to have the published results incorporated into the nuclear databases. However beyond that final step each of the databases are intimately intertwined and absolutely essential for each of the processes indicated; from planning experiments and writing proposals, to executing experiments, publishing results and refereeing publications. Some examples are described in more detail below.

The important role that evaluated nuclear data play in the execution of an experiment was especially highlighted for FRIB, where any radioactive beam produced via fragmentation is unlikely to be pure. Evaluated data then become critical to identify, in real time, particles being delivered to an experiment. Figure 4 shows as an example a recent FRIB experiment utilizing a $^{37}$Al beam. While the element can be determined from an energy loss measurement there are ambiguities with the respect to the isotope. Measuring γ-ray spectra gated on individual isotopes make a unique identification possible. However, this relies on the knowledge of the γ-ray transitions of the β-decay daughters. In the γ-ray spectrum in the figure, known transitions for $^{37}$Si as listed in ENSDF are clearly visible. This can be taken a step further, in that it is possible to query the databases for a specific γ-ray transition energy and generate a listing of nuclei to which the γ ray might belong. For ENSDF to be useful in this respect, it must be kept up to date. Results from new radioactive beam facilities need to be quickly and carefully incorporated into the databases, so that subsequent experiments can benefit from their results.

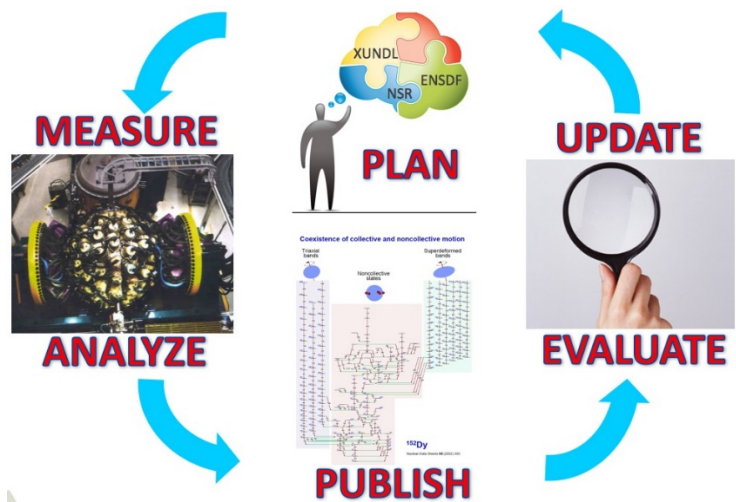

*Figure 3: Schematic representation of the life cycle of a nuclear physics experiment (Figure presented by M. Carpenter).*

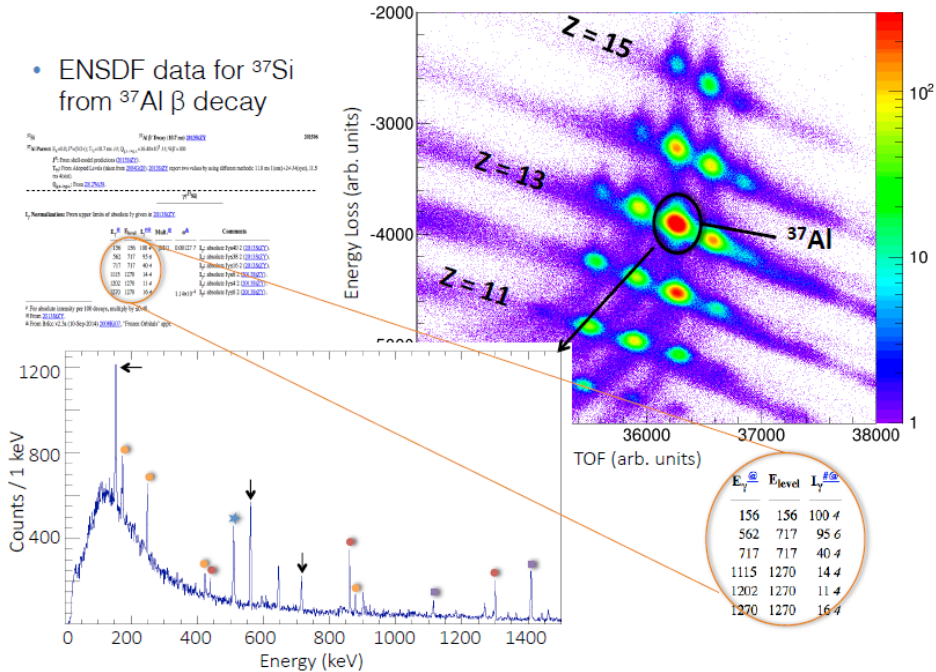

*Figure 4: Example of particle identification using the known γ-ray transitions in $^{37}$Si following the β-decay of $^{37}$Al (Figure presented by H. Crawford).*

When interpreting the data, it often is useful to consider not only the nucleus under study, but the neighboring nuclei and their properties. ENSDF and XUNDL contain a wealth of information which, when displayed in graphical form, can be used to explore correlations in data and to benchmark models. Investigating data through different perspectives yield different insights and understanding into the underlying nuclear structure. As an example, a common observable used to gauge the amount of deformation in nuclei is the ratio of the $4^+$ to $2^+$ transition energies, the so-called R4/2 ratio. This is plotted in Figure 5 (top) using an application currently available to users through the NNDC website. While some trends are evident, like increasing R4/2 values with increasing neutron number, the fine details of the evolution of this quantity are not immediately obvious from the plot. The bottom panel of Figure 5 shows the same quantity now plotted as a function of proton and neutron number. The large gap in the right plot indicates the presence of a sub-shell closure, which wouldn't be discovered if only the plots in the top and left panels were analyzed. This example demonstrates that the capability of plotting evaluated quantities using ENSDF/XUNDL data and flexibility in the plotting such that any evaluated quantity could be easily selected would be extremely useful for enhancing nuclear structure research. It also would be beneficial to be able to generate level schemes directly from ENSDF/XUNDL.

Within the evaluation process itself, there is room for improvement, with some portions of the ENSDF/XUNDL format potentially causing confusion for the general user. While the detailed nomenclatures of ENSDF are described in the policy section of the Nuclear Data Sheets journal, typically these are not consulted by the average user. Thus, the data community needs to make attempts to improve the presentation and clarity of data. Particular examples include conversion coefficients, use of upper and lower limits which may lead to unphysical values, and notations for multipolarities. A first step in this direction is the new JAVA-NDS format which significantly improves the transparency of the Nuclear Data Sheets journal. These types of changes should also be incorporated into the web-based retrieval system. The web based retrieval system should develop clear help sections and quick reference guides which users can consult for definitions of nomenclature, policies and units.

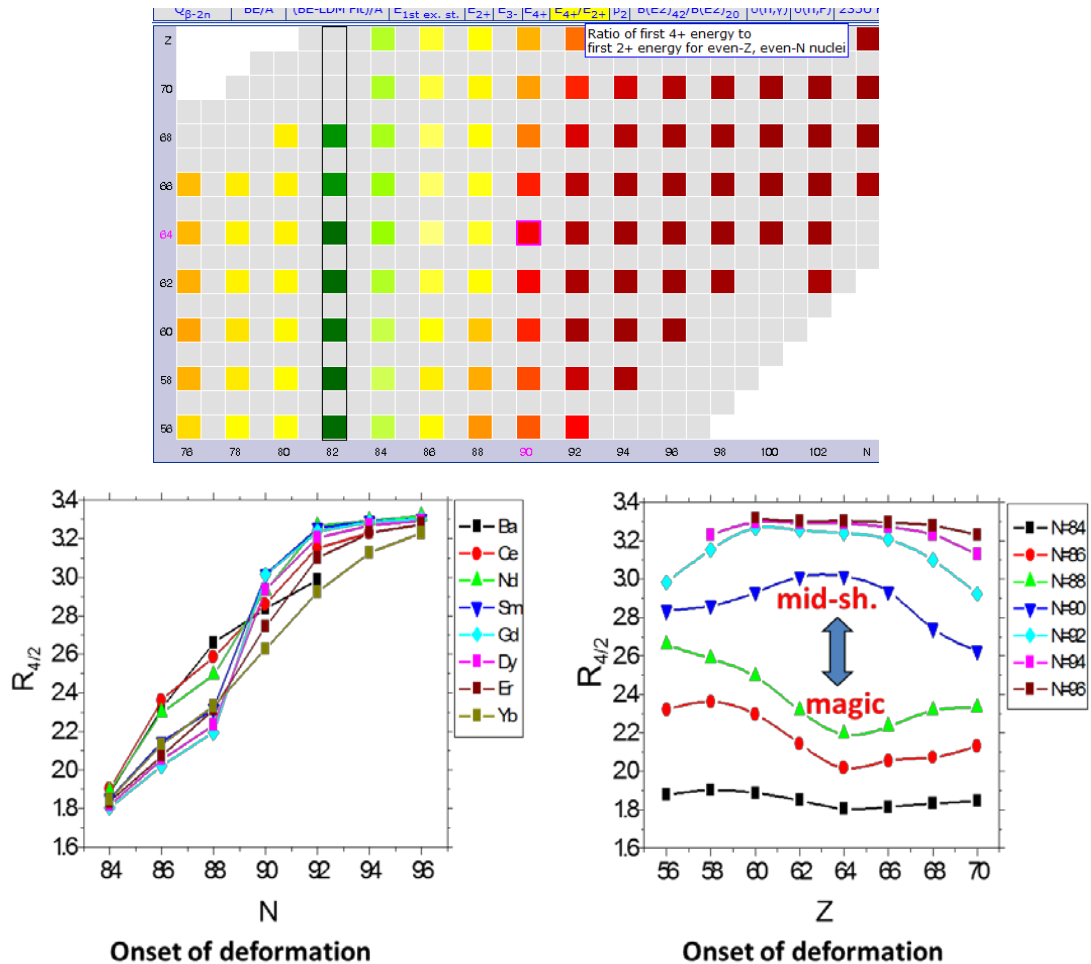

*Figure 5: (Top) $R_{4/2}$ as a function of proton and neutron number as generated by the NuDat program. Colors indicate the magnitude of the quantity, green being smaller and red being larger. (Bottom) $R_{4/2}$ as a function of neutron number for different elements (left) and as a function of proton numbers for different isotopes (right) (Figure presented by R. Casten).*

In order to maximize the impact of the USNDP services and databases it would be beneficial to improve the communication and interaction with the community. For example, apparently the existence of the XUNDL database is not well known among research scientists. Targeted seminars describing all databases and applications available to the community by the data evaluators at major facilities and conferences should be encouraged.

Typically, the evaluation process is performed in mass chain units, meaning each nuclide of a given mass number, A, is evaluated at the same time and published together. This is a convenient unit for the evaluation procedure, as the majority of radioactive decay follows along the same A line. However, this is not the path most experimental endeavors follow. For example. Figure 6 highlights the regions of the nuclear chart which were studied in recent campaigns of the new GRETINA array. Horizontal or topical evaluations are a much better match to the way experimentalists think about the nuclear chart, and would provide a valuable tool for planning experiments and interpreting structure in different regions. Some areas of recent experimental interest include the island of inversion, N=Z nuclei, superheavy elements and β-delayed neutron emitters. For such topical work, open discussions between the data evaluators and the experts in the community would help to strengthen the quality of the evaluation.

The information compiled and evaluated in ENSDF and XUNDL needs to evolve with current experimental research interests. One example is continuum data, such as measurements of radiative strength functions, β-delayed neutron spectra and data coming from new Total Absorption Spectroscopy measurements. The latter, is useful in a wide range of fields including nuclear structure but also reactor kinematics and antineutrino spectra calculations. Such additions could potential require expansions or changes to the current ENSDF format.

The current version of ENSDF contains purely experimental nuclear structure data. To fully exploit the information in ENSDF, the data community developed many tools for querying, manipulating and plotting the data. As the breadth and scope of theoretical calculations increases, it would be desirable to take advantage of such tools to explore theoretical calculations which are capable of providing predictions for the entire nuclear chart. One example would be to create theoretical analogs of ENSDF or its graphical interface, NuDat. Tools to tie together and compare the experimental and theoretical results would further enhance both efforts.

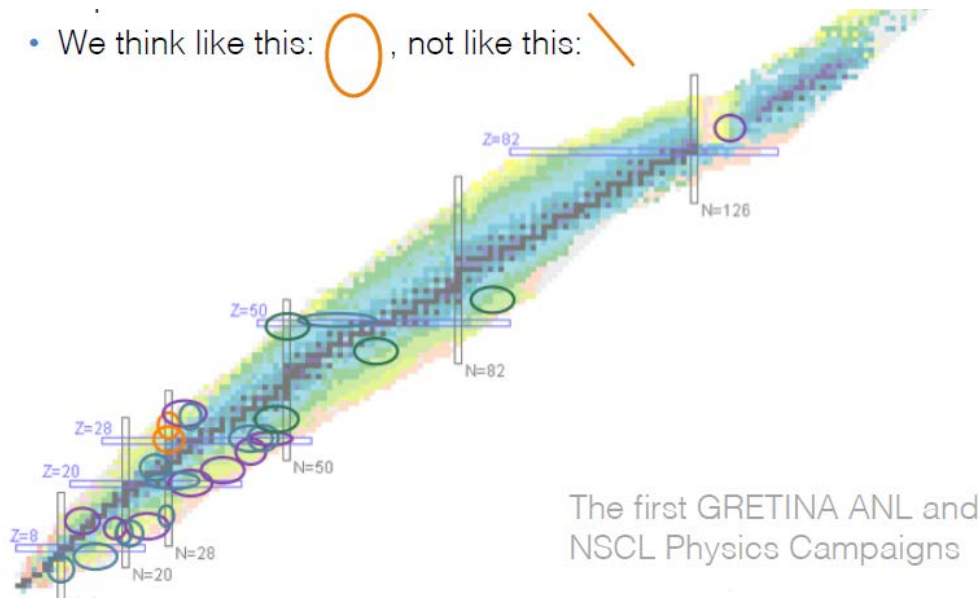

*Figure 6: Nuclear chart indicating the regions of study probed by the recent GRETINA campaigns (Figure presented by H. Crawford).*

Finally, as mentioned above, fragmentation reactions can produce a number of nuclei, beyond the ones which are the primary focus of the experiment. Looking to the future, and specifically FRIB, all of the data produced in these experiments, not just those that make the "headlines", should be incorporated into the databases and disseminated back into the user community. The data community should begin dialog with the large facilities and work to generate data archiving schemes, so that no data is left behind.

# 2. Nuclear Reactions

Nuclear reactions and nuclear structure are tightly connected. On the one hand, nuclear structure properties (e.g., *J, π, $E_x$, $T_{1/2}$, decay branches*) are important ingredients for nuclear reactions measurements and calculations. On the other hand, many structure properties are measured and deduced in reactions, thereby making a good understanding of the reaction mechanism necessary in order to properly interpret the observables. The workshop highlighted some of these interdependencies and emphasized the need for properly compiling and evaluating all of the necessary physical observables, including those that are not described using the quantities represented in ENSDF (mostly $J^\pi$ and $E_x$) or ENDF (mostly cross section).

The importance of reliable nuclear reaction data for the applications community has already been stressed in the recent white paper of the Berkeley workshop on Nuclear Data Needs and Capabilities for Applications. Thus we concentrate here on the significance of understanding nuclear reactions for basic nuclear structure physics and nuclear astrophysics.

Cross section measurements of transfer reactions are essential for elucidation of the single-particle properties of nuclei. Although the structure of stable nuclides is in general well understood, the knowledge of single particle occupancies and vacancies is critical for double β-decay studies and for the exploration of the tensor force towards the driplines.

Most the transfer reactions with stable beams in forward kinematics were performed about fifty years ago and the data were analyzed with the limited codes of the time. In contrast, the present (and future) radioactive beam experiments are performed in inverse kinematics and at potentially different beam energies. Thus, it is important to validate these new data by comparing them to the normal kinematics measurements. Unfortunately the raw data from these experiments were not preserved in easily accessible form, so that they are difficult to compare to the modern reaction model codes.

In many cases, only the derived spectroscopic factors were listed in tables and the cross-section data (including angular distributions) were only shown in graphical form. Therefore, it would be very beneficial to digitize and compile these historical data. This is probably the only practical way to recover these data, as most of the accelerators and spectrometers used to gather these high precision data do not longer exist.

In addition to the spectroscopic factor measurements, transfer reactions in inverse kinematics with radioactive beams are also used to extract resonance parameters of particle-unbound states at and beyond the driplines. These states are typically reconstructed from invariant mass or missing mass measurements. In order to extract the resonance energies and width the data have to be compared to simulations that are based on reaction models. For broad resonances, the decay times are comparable to the reaction time scales so that the direct interplay between the reaction and the involved resonances has to be taken into account.

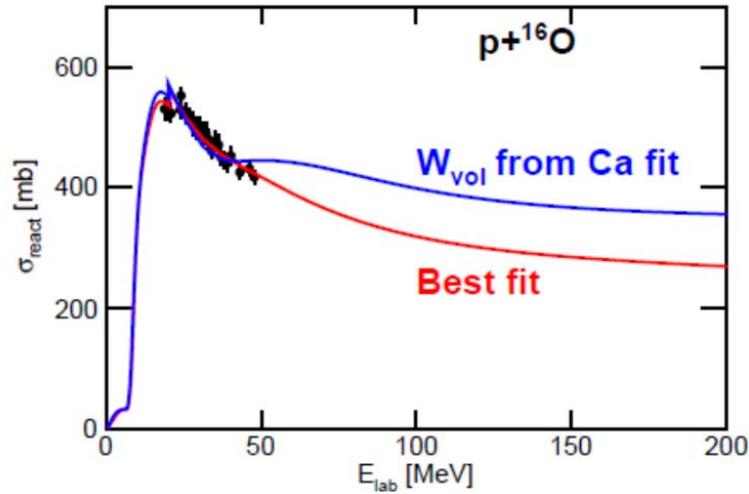

*Figure 7: Measured p+$^{16}$O reaction cross sections (up to 50 MeV) compared calculations using an extrapolation from the best fit to the data and calculations based on calcium data (Figure presented by L. Sobotka).*

In general, a more consistent analysis of these transfer reaction would benefit from an online repository of reaction codes and global optical model potentials (OMP). In fact, the OMP are not well determined as one would expect and surprisingly few data exist. Specifically, total reaction cross section data for neutron and proton induced reaction on isotopically separated targets are scarce. As an example, Figure 7 shows the reaction cross section for protons on $^{16}$O compared to two different fits. There are no data beyond 50 MeV and it also should be mentioned that there are no data at all for p+$^{18}$O.

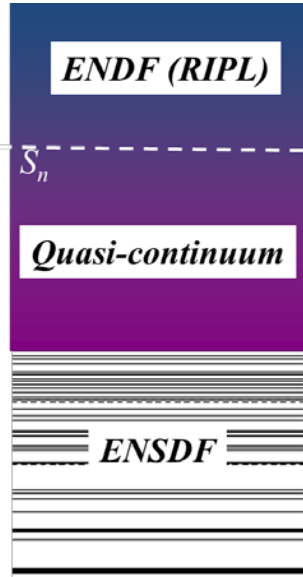

*Figure 8: Nuclear levels and level density as a function of excitation energy. In stable isotopes the neutron separation energy (Sn) is in the region of high level densities (~8 MeV) (Figure presented by L. Bernstein).*

The interplay between reactions and structure is shown in Figure 8, where the level density as a function of excitation energy in a nucleus is shown. While at low energies individual levels can be identified (and measured), at high energies the number of levels becomes so large that they can be best described as a "quasi-continuous" level density (LD) since the levels involved are still bound. The individual levels of a nucleus contain detailed information about the nuclear structure as described in the previous section and their properties are compiled and evaluated within ENSDF. The high energy region is important for the modeling of reactions for applications, such as neutron capture reactions on stable isotopes, which populate states at excitation energies (~8 MeV) where the LD is high. An interesting region is the transition from individual levels to high LD. In the quasi-continuum (QC) different γ-ray de-excitation channels can have a significant impact on the final reaction products, thus it is critical to know the LD and the related γ-ray strength functions (GSF). Similarly, QC levels also provide insight into the nature of emergent collective phenomena, such as the pygmy electric dipole and magnetic spin-flip resonances. There is recent evidence of a systematic enhancement of the GSF indicative of the presence of pygmy E1 and/or M1 collective motion.

QC properties are not easily evaluated as they are continuous functions. Furthermore, there is often significant disagreement between experimental results, since the determination of the LD and GSF involve the use of modeling and simulations, thus making impartial evaluations essential. The knowledge of these functions is crucial for nuclear astrophysics as the relevant nuclear capture reaction take place predominantly in neutron-rich nuclei where the neutron separation energy is significantly lower. Thus, the γ-ray strength functions determine the path that r-process nuclides decay back to the stable isotopes. For all of these reasons it is clear that new evaluation methodologies need to be developed for quasi-continuum properties that includes both a treatment of the recommended values and a good representation of the attendant uncertainties.

This is just one example where improvements in the experimental techniques allow for significantly more detailed and exclusive measurements of observables that have to be compiled and evaluated. Another example is the correlated data collected with the DANCE detector at the Lujan Center of LANSCE. Prompt γ-ray spectra and multiplicity distributions from neutron induced fission of uranium and plutonium were measured in coincidences with fission fragments in the PPAC detector, as shown in Figure 9.

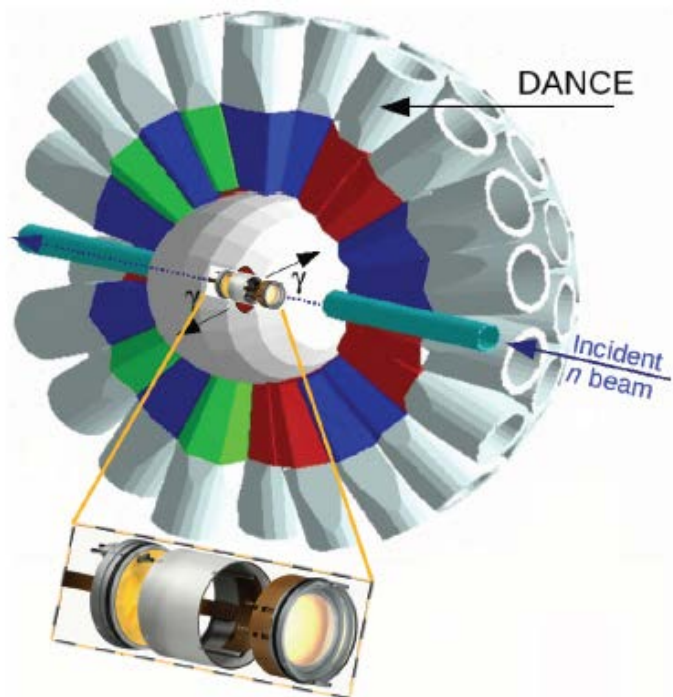

*Figure 9: Schematic diagram of DANCE including the PPACs for coincident fission fragment detection (Figure presented by C. Wu).*

The total γ-ray energy is a function of the γ-ray multiplicity and, as a consequence, the extracted average total γ-ray energy depends on the correlation between those two quantities. These observables can also be calculated by theoretical fission models. Therefore, it is important to document and compile these measured correlations, so that they can be compared to the results from simulations.

Monte-Carlo simulation codes that would benefit from access to correlated data are for example GEANT4 and LISE++. In general, these codes rely on easy accessible nuclear data libraries. GEANT4 is a toolkit for Monte Carlo simulation of the passage of particles through matter. The package G4LEND models low energy nuclear interaction to reproduce cross section and final states of reactions. Clearly, the physics performance of G4LEND depends on the quality of the libraries and the underlying data.

LISE++ simulates the production and separation of isotopes and is the primary tool to predict the intensities and purities of radioactive beams at FRIB and other radioactive beam facilities. Several different reaction mechanisms can be modelled including projectile fragmentation, fusion-evaporation, fusion-fission, Coulomb fission, and abrasion-fission.

The accurate predictions of radioactive beam intensities is critical as a factor of two can determine if a proposed experiment is feasible or not. This accuracy is challenging to achieve. For example, the production cross sections of neutron-rich isotopes for a given element from projectile fragmentation drop exponentially towards the neutron-dripline.

LISE++ includes databases of a variety of nuclear properties, such as atomic masses, isomeric states, fission barriers, experimental production cross sections and decay branching ratios.

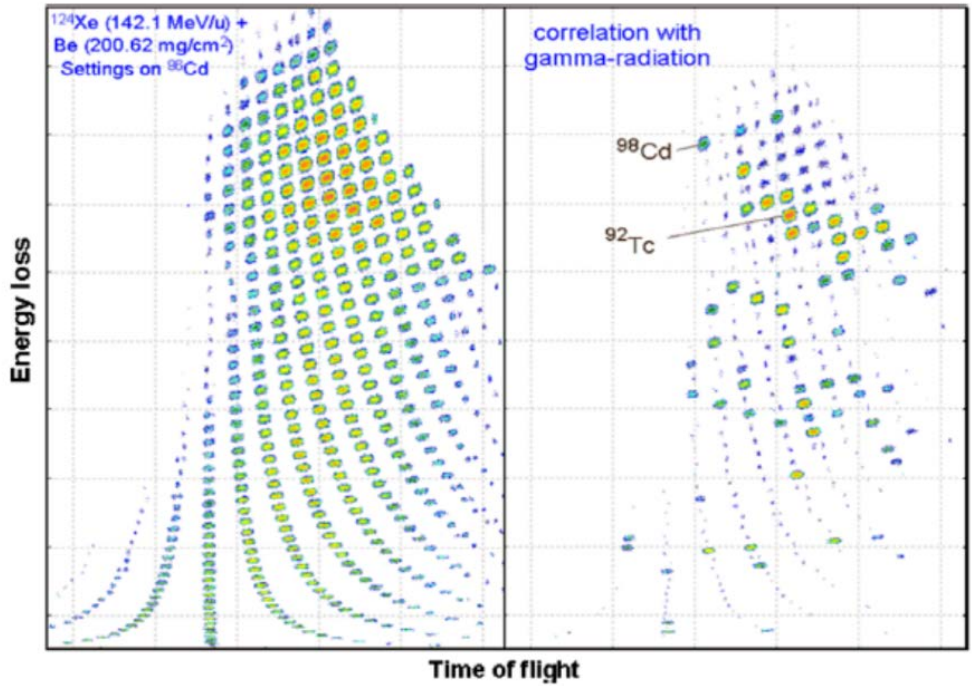

*Figure 10: Results of LISE++ simulations displaying all nuclides produced in the reaction $^{124}$Xe + Be (left panel) and those in coincidence with γ rays (right panel) (Figure presented by O. Tarasov).*

The importance of up-to-date knowledge of isomeric states for the design, planning and preparation of a radioactive beam experiment is shown in Figure 10. The particle identification plot in the left panel shows all nuclides produced in the reaction $^{124}$Xe + Be as simulated by LISE++. In order to select the desired beam for a given experiment, it is crucial to correctly identify the correct nuclide of interest. Measurements of γ rays in coincidence with the detected nuclides can reveal isotopes with μs-isomeric states, which in turn can be then compared to known isomers in a given mass region, as shown in the left panel of the figure. Thus, up-to-date database that contains the properties of isomeric states can become a powerful particle identification tool.

# 3. Nuclear Astrophysics

Nuclear Astrophysics is focused on determining the cosmic origin of the elements and on elucidating the nuclear physics phenomena that drive the evolution of universe and the explosion of stars. This interdisciplinary field has extensive nuclear reaction and structure data needs, which combined with specialized data processing steps provide critical input for simulations of various cosmic systems. USNDP activities should be expanded to include efforts specifically targeted for nuclear astrophysics in order to maximize the scientific return on recent investments in this area. A detailed plan for developing effective interactions between the nuclear astrophysics and USNDP is outlined in Appendix 2.

Nuclear astrophysics involves studies of some of the most fascinating systems in the Universe, from explosions like novae and X-ray bursts to the properties of neutron stars and their mergers, from the earliest stars in the universe millions of times more massive than the Sun to stars forming today, from Red Giants to White Dwarfs to Black Holes, from our Sun to the most distant stars. Such studies are a crucial component of the field of nuclear physics, addressing one of the four overarching questions in nuclear science from the National Research Council's 2013 Assessment of Nuclear Physics and echoed in the 2015 Long Range Plan for Nuclear Science: "How did visible matter come into being and how does it evolve?" Determining the cosmic origin of the elements, and obtaining a better understanding of the nuclear physics that drives stellar evolution and stellar explosions, are the primary focus areas of this research. The importance of this work is confirmed in its role as a major motivation for the top priority construction project in U.S. nuclear physics, the Facility for Rare Isotope Beams (FRIB).

One of the most compelling aspects of this research is the inherent linking of the physics at the tiny distance scales of subatomic nuclei to stellar phenomena at scales $10^{24}$ times larger. Such a linkage requires extensive and diverse nuclear data sets that serve as the foundation for simulations of nucleosynthesis, stellar evolution, and stellar explosions. Predictions made in such simulations are, in many cases, extremely sensitive to the input nuclear data. Figure 11 indicates the relationship of some of the thermonuclear burning processes in the cosmos to the relevant nuclides in the nuclear chart.

The data needs for nuclear astrophysics cut across the traditional boundaries of nuclear structure and nuclear reactions. Furthermore, a specific set of processing steps is required wherein nuclear data is combined with nuclear theory and astrophysical theory to produce thermonuclear reaction rates. These rates are crucial, providing the foundation for all nucleosynthesis simulations. Without this specialized processing (e.g., evaluation, extrapolation with theory, conversion to reaction rates), the latest experimental results and global nuclear theory calculations cannot be used in astrophysical simulations. The rapid flow of information between nuclear experimentalists and theorists, nuclear data evaluators, and astrophysics modelers is therefore essential for progress in this field.

Reaction cross section measurements that are of specific interest include proton and neutron induced reactions on stable, as well as on neutron- and proton-rich, nuclides up to energies of a few MeV. Photo nuclear reaction cross sections are also of interest for stable and proton-rich nuclides above iron.

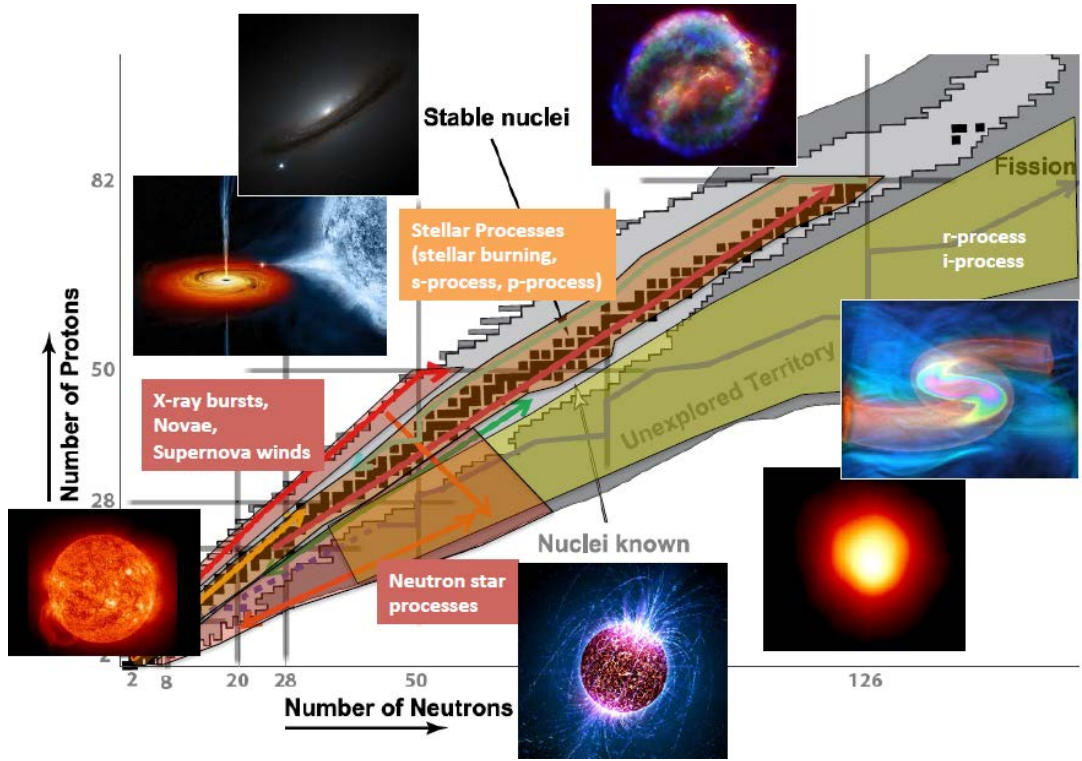

*Figure 11: Chart of Nuclides indicating the regions of the various thermonuclear burning processes in the cosmos (Figure courtesy of H. Schatz).*

Detailed nuclear structure information for a very broad range of stable and unstable nuclides are needed for accurate astrophysics simulations. This includes nuclear masses, properties of low-lying excited states (excitation energy, spectroscopic factor, spin, parity, partial and total decay width, resonance strength, gamma-ray strength functions, level densities, etc.), weak interaction rates and fission barriers.

Since it is not practical to measure these properties for all nuclides, the thermonuclear reaction codes have to rely on theoretical modeling to derive global trends. For example, the shell model is needed to calculated missing resonance data, R-matrix analysis is required to extrapolate to low temperatures, and the statistical model is essential to extrapolate to higher temperatures.

Thermonuclear reaction rates are the foundation of most nuclear astrophysical simulations. For stable nuclei, a large numebr are determined by thermally averaging cross sections. Many others are based on analytical calculations using the properties of known and predicted levels near particle thresholds. The processing steps needed to generate reaction rates are specific for each reaction type. To use these rates in astrophysical simulations, however, it is essential to have complete coverage of rates across the nuclide chart. We note that while thermonuclear rates from ENDF/B or related application libraries have been determined, their utilization in astrophysics simulations is problematic because they lack reactions on unstable nuclei, and are hence incomplete, and because the relevant cross sections are often optimized at energies far above those of interest to astrophysics.

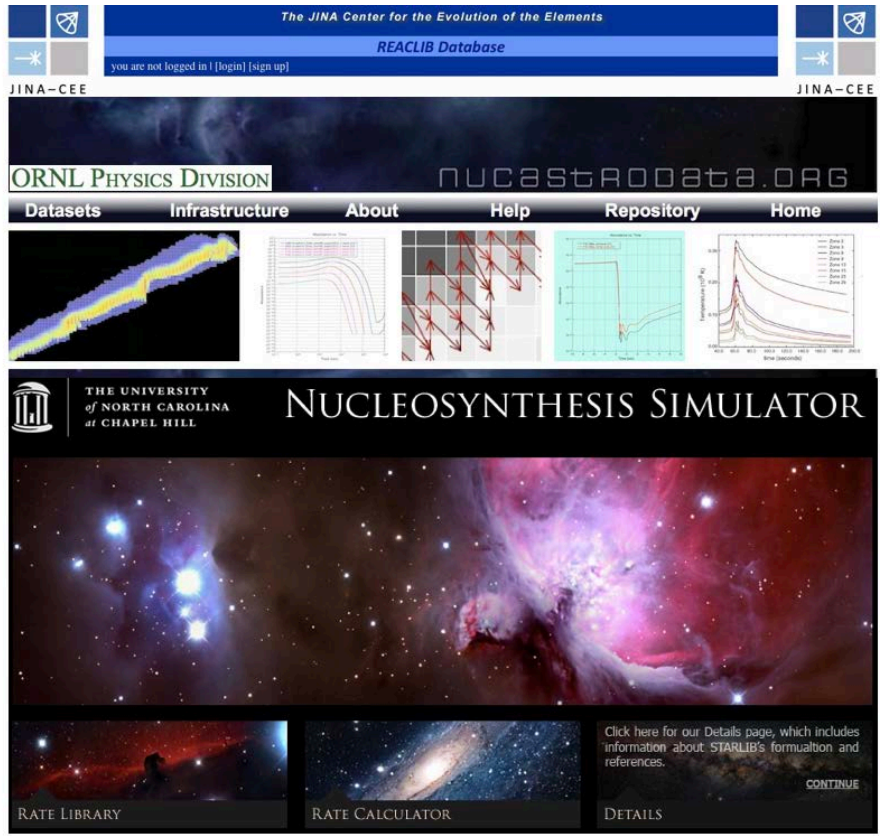

*Figure 12: Examples of nuclear reaction rate libraries.*

In addition, the thermonuclear reaction rates have to take into account the stellar environments (enhancement), where a fraction of nuclei are thermally excited into low-lying states. Thus, the ground state cross sections measured in the laboratory have to be corrected for contributions from possible reactions on these excited states.

Since there are typically hundreds to thousands of thermonuclear reactions relevant for studies of any given astrophysical environment, the collection of rates into libraries is essential for progress in the field. The development of comprehensive libraries for thermonuclear reaction rates should be strongly coordinated with the above mentioned resources and other existing databases. Figure 12 shows a few examples of existing libraries.

The nuclear astrophysics community has advocated for the uniform adoption of the JINA REACLIB library as the standard for nuclear astrophysics simulations. It will also include additional smaller-scale libraries, created by different research groups that will be distributed to the community. Strong support for continuing to grow this library, as well as the associated software tools used to manage, modify, and share the rates, is essential for progress in the field.

# 4. Nuclear Theory

One long-term goal of nuclear theory is to achieve a comprehensive and unified description of all nuclei and their reactions based on the fundamental interactions between the constituent protons and neutrons. Calculations and predictions of properties for specific nuclides can be directly compared with the latest experimental results or with the evaluated data for these nuclides in ENSDF.

However, with the continued increase in computational power it is possible to generate large scale theoretical data sets for broad ranges of nuclides which have to be compared to the available evaluated data. Figure 13 shows a schematic of the interplay between experiment, theory and nuclear data that can lead to new discoveries.

"Horizontal" evaluations for basic nuclear properties (for examples masses, B(E2) values, quadrupole and magnetic moments, etc.) should be easily accessible to be compared directly with the corresponding theoretical values. Some initial efforts along these directions exist already. For example, in Figure 14 nuclear masses calculated with the FRDM1995 model is compared with the AME2012 evaluation for all nuclide across the nuclear chart. Another example is the Mass Explorer website (massexplorer.frib.msu.edu) that contains results from large-scale Density Functional Theory calculations of ground state properties of even-even nuclei throughout the nuclear landscape.

In the long-term, it could be envisioned to interface theoretical and evaluated databases to compare theory and experiment interactively. Both data sets should always include the relevant uncertainties, covariances and other relevant information on how the data were generated. It will be important that any merged databases consisting experimental as well as theoretical data clearly distinguish between the two different input sources.

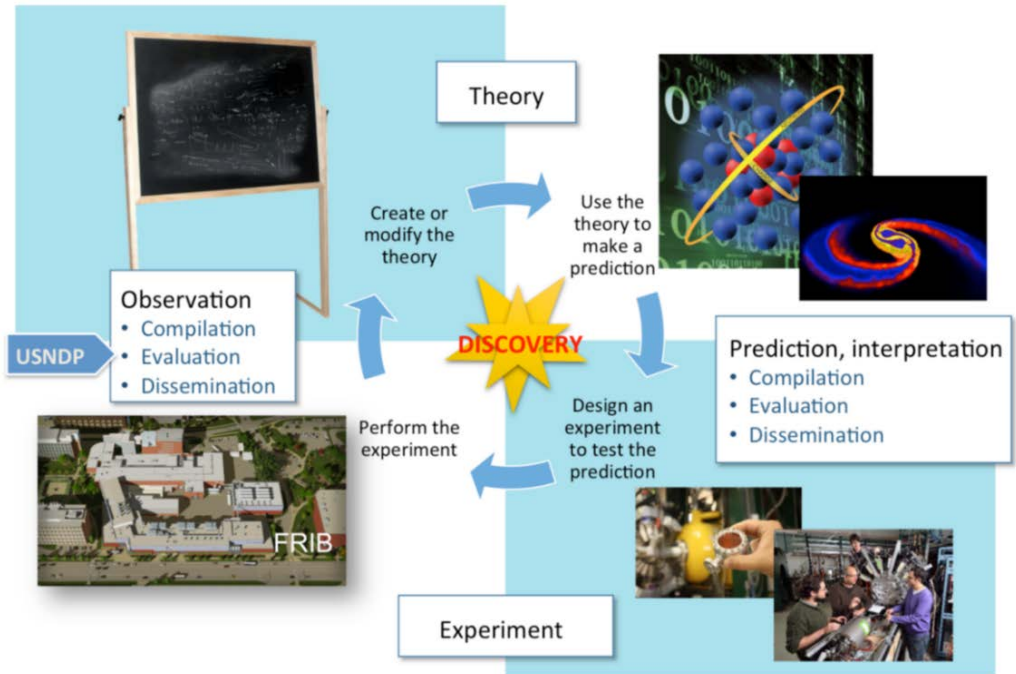

*Figure 13: Interplay between experiment, theory and nuclear data leading to new discoveries (Figure courtesy of W. Nazarewicz).*

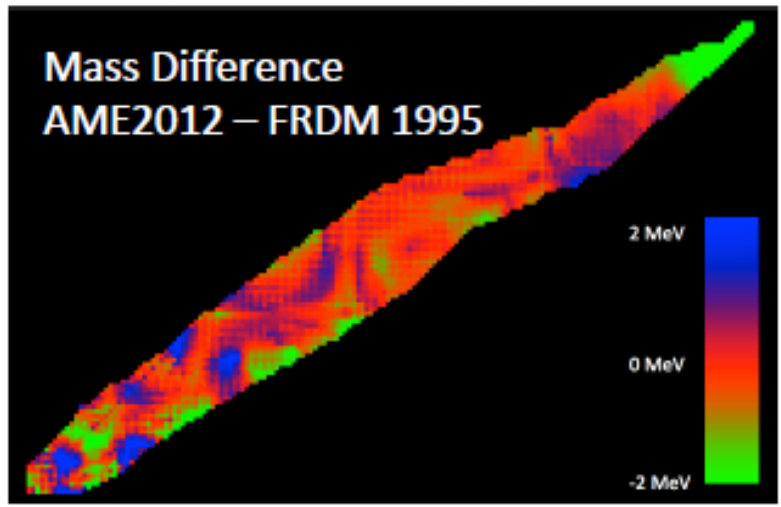

*Figure 14: Chart of nuclide showing the mass differences between the AME2012 evaluation and the results from Finite Range Droplet Model (FRDM2012) calculations. The plot was generated using software available at nuclearmasses.org (Figure presented by M. Smith).*

It probably should also be considered to systematically record and document the computer codes used to generate the theoretical data along with typical or systematic input and output files. This will require good cooperation with the code authors, so that version numbers or repository archive information is given to assist in the reproducibility of replication efforts. Ideally, the codes used in published works could be saved along with full details for replicating its results.

The following examples reiterate the importance for close relationships between theory and nuclear data, albeit some of them have already been discussed in the previous sections on nuclear structure, reactions and astrophysics.

The properties of light nuclides are described by *ab initio* nuclear calculations. Even the most microscopic method for nucleonic systems start from a nucleon-nucleon reaction designed to fit the world set of scattering data for nucleons scattering on nucleons. A compilation, evaluation or validation of such data is thus the essential starting point for the *ab initio* project for light nuclei, whether for their structure or for reactions involving them. The NN-online database at http://nn-online.org/ for many years taken this role.

For heavier nuclides, structure calculations such as using the shell model, proceed by fitting various nuclear matrix elements to a set of known levels in a number of representative nuclei across the nuclide range of interest. Thus it is imperative that the nuclear data evaluations contain up-to-date information of the energy, spin and parity of the levels, in order that they may with their theory counterparts. Figure 15 shows as an example the results of shell model calculations for $^{53}$Fe with the corresponding evaluated data set extracted from ENSDF.

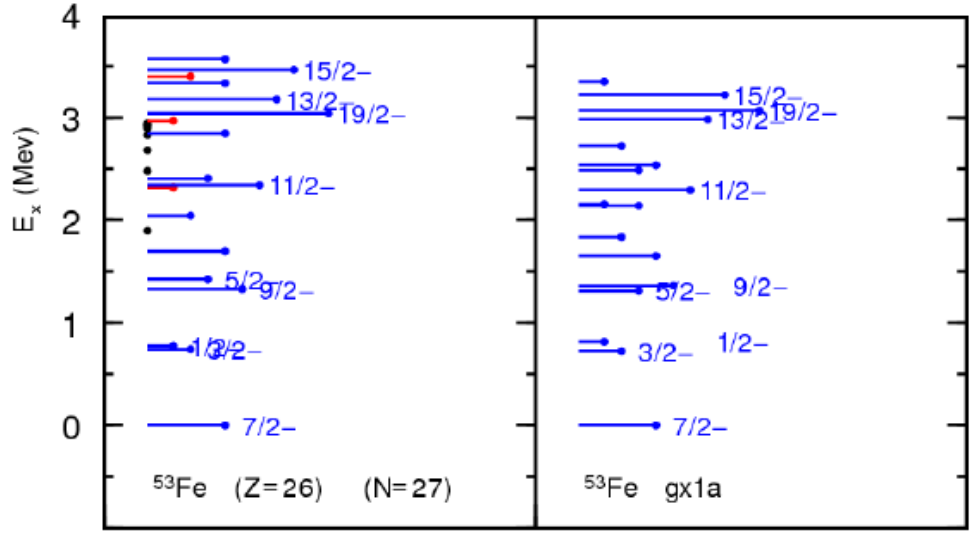

*Figure 15: Level scheme for $^{53}$Fe extracted from ENSDF (left) compared to shell model calculations using the gx1a interaction. The length of the bars indicates the magnitude of the spin (Figure presented by B.A. Brown).*

Nuclear structure data is also essential for astrophysics. Mostly it focuses on ground-state properties such as masses, charge radii, and half-lives, although information about radial densities, low-lying excited states, giant resonances and photon strength functions are often useful. Since many of the needed properties are not yet nor will not ever be measured, the experimental databases have to be supplemented by theoretical predictions as alluded above.

Cross sections of direct nuclear reactions are compiled and validated within the EXFOR database. Traditionally this has included angle-dependent and energy-dependent distributions, as well as total, capture and fission cross-sections. In recent decades breakup cross sections have also been measured, and recording and evaluating continuous breakup data requires detailed knowledge of detector geometries and kinematic reference frames.

As mentioned earlier, continuous data are in general more difficult to evaluate. In all cases, it is important to compile and record the original (raw) data before any model-dependent analyses are performed. For example for neutron-induced reactions the measured γ-ray distributions should be archived together with the derived photon strength functions. This will make it possible to reanalyze the data in the future in cases where more sophisticated theoretical models become available.

Of course the derived data and reaction model parameters needed to fit the data are also important to compile. These include optical potentials (local, regional, or global), spectroscopic factors, deformations, asymptotic normalization coefficients (ANCs), and R-matrix parameters. A global set of parameters is necessary for applying the models to predict and calculate cross sections of important reactions which are not experimentally accessible.

# 5. Neutrinos Science and Fundamental Symmetries

The relationship between neutrino and fundamental symmetries and nuclear data is different from the nuclear structure, reactions, and astrophysics discussed in the previous sections. Although they are fundamental sciences they rely on high quality and high precision nuclear data similar to the applications covered by the Berkeley workshop and subsequent white paper.

There are two different broad needs. The first arises from practical challenges that emerged during the preparation, execution or analysis of the experiments designed to address certain fundamental questions. Two examples which are discussed below are the detailed knowledge of the neutrino flux in reactors and the accurate calculation of the environmental background in neutrinoless double β decay (0υββ).

The second need for nuclear data is directly related to the interpretation of measured data. The nucleus can serve as a laboratory to study some of the present day questions in neutrino science and fundamental symmetries. However, it requires the detailed knowledge of a variety of nuclear structure properties for a range of different nuclide. Thus a comprehensive compilation and evaluation of these properties is essential. Below, three examples are described; the determination of the neutrino mass, the test of the unitarity of the CKM (Cabibbo-Kobayashi-Maskawa) matrix, and the search for the Electric Dipole Moment.

Recent studies antineutrinos from reactors have provided evidence the oscillation between different neutrino flavors. However, the data reveal two characteristics of the measured neutrino spectrum that are not well understood. The first is an apparent downward shift in the neutrino flux measured at distances close to the reactor, the so called "anomaly", which describes that only 95% of the expected flux is observed. The second is an apparent "bump" in the measured antineutrino energy spectrum near E~6 MeV, as indicated in Figure 16. The findings are based on a comparison of the measured spectra with rigorous calculations based on expected fission reaction dynamics of $^{235}$U, $^{239}$Pu, $^{241}$Pu and $^{238}$U. If the anomaly and bump are found to be valid features of the reactor antineutrino spectrum and flux, they could indicate new physics, such as existence of a fourth neutrino species. The neutrino yields are expected to correlate predictably with the power output from the reactors, and they are rooted in Monte Carlo simulations that follow the reaction chain that describes the path that fission products follow as they β decay toward the stability line. In such a model, accurate nuclear data are required at each step along the way. In addition to fission cross sections and fission product yields the simulations rely on complete and precise evaluations of β-decay feeding intensities and their uncertainties, together with data on β-delayed particle emissions.

The case of the anomaly and bump in the reactor antineutrino spectrum highlights a few critical points. Firstly, the analysis and interpretation of experimental data relies on sophisticated Monte Carlo simulations that use a variety of nuclear structure and decay data. Secondly, there is a need to solve inaccuracies of the data that may be rooted in experimental errors or simple shortcomings of the experimental techniques that inhibit a full interpretation of the complete physical picture. In this context, high-quality data collected via discrete γ-ray spectroscopy and the total absorption γ-ray spectrometry (TAGS) techniques are needed to improve the knowledge. The complete evaluation of these data and their prompt inclusion into the ENSDF database is needed. While systematic uncertainties are important in the relationship between TAGS observables and traditional β-decay analyses, the two approaches must be regarded as complementary rather than competitive; congruency must be found between both approaches. The critical issue is that the analysis of the reactor antineutrino spectra relies on complete and accurate evaluation of nuclear data, including fission product yields and nuclear structure and decay properties.

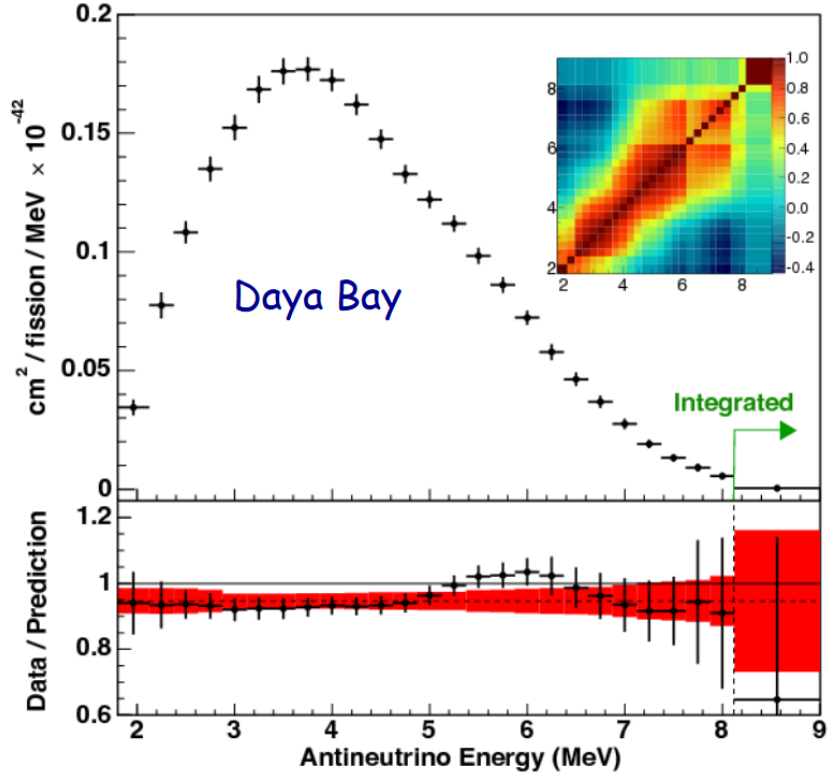

*Figure 16: Antineutrino spectra measured by the Daya Bay collaboration that place strong constraints on the nuclear databases (Figure presented by B. Balantekin).*

In addition to the precise understanding of the fission process itself, the reactor antineutrino spectrum is influenced by environmental variables, such as fuel-rod claddings or nuclear materials that may absorb neutrons and slow the fission process. Reactor poisons, such as $^{135}$Xe are well known to influence reactor performance. The production or presence of any neutron absorbing poison, along with the associated cross sections relevant to the influence these materials may have for slowing down the fission process must be fully accounted for in the Monte Carlo simulations.

The second example for the importance of environmental parameters is the search for 0υββ. It has been posturized that the neutrino is its own antineutrino since 1937; however, proof is elusive even in modern times. The primary approach for validating this hypothesis has centered on the observation of 0νββ. On a fundamental level, the efforts for observing 0νββ decay have utilized detailed analysis of nuclear data to evaluate spectroscopic information to determine the optimal configurations that may permit such an observation. In addition, on a practical level, the expected rate for such a rare process has moved these searches into locations where the background surrounding the experiment must be as low as possible and must be well characterized. The experiments are buried deep underground and are surrounded by radio purified materials to minimize background counts in the region of interest that could mask the results or worse, give rise to false affirmative results. The present programs searching for 0νββ decay have developed state-of-the-art Monte Carlo simulations of reactions on radioactive materials that cannot be removed from the environment or from atmospheric processes that reach deep into the earth's surface. Reliable interpretation of these low-event rate studies relies on nuclear reaction data and nuclear structure data to provide accurate details on their environmental backgrounds so that dependable results can be obtained in their analyses.

The observations of neutrino oscillations mentioned above implies that neutrinos have mass. Traditional studies of the neutrino mass have focused on the analysis of anomalies in the phase-space spectrum of the 3-body neutrino emission near the Q-value energy limit. In addition to precise determination of the shape of the β spectrum near the endpoint, these analyses also rely on accurate determination of nuclear masses and β-decay Q-values.

Especially interesting could be β-decay transitions with ultra-low (UL) energies as indicated in Figure 17. If cases can be identified, where the parent ground state energy is nearly degenerate with a level in the daughter nucleus that is allowed, then observation of the β-delayed γ-ray transition from that state can set a limit on the neutrino mass. Examples where the degeneracy is within 10 keV include the parent nuclides $^{72}$As, $^{72}$Se, $^{75}$Se, $^{88}$Zr, $^{96}$Tc, $^{112}$Ag, $^{113}$Ag, $^{115}$Cd, $^{156}$Eu, $^{159}$Gd, $^{186}$Re, $^{188}$W, and $^{199}$Hg. The meaningfulness of such observations depends directly on the precision of the nuclear data for nuclear masses, Q-values and evaluated level energies. New limits could only be set with precisions in the nuclear data that are on the order of 0.5 keV or better for all related quantities.

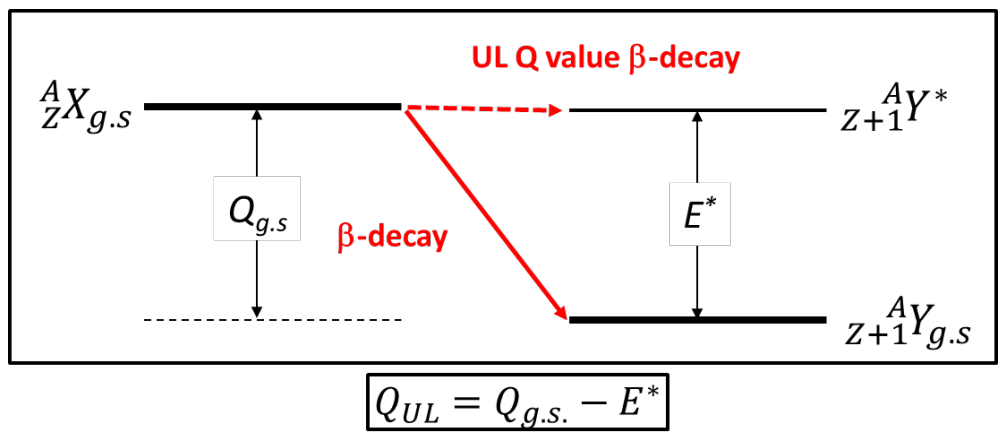

Figure 17: Decay scheme for the β-decay with ultra-low (UL) energy Q-values (Figure presented by M. Redshaw).

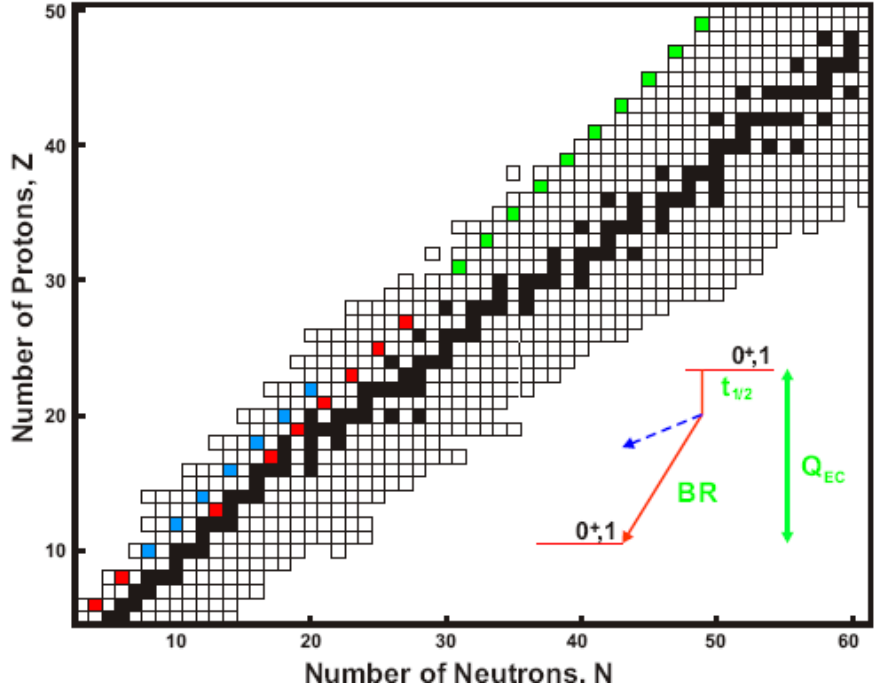

*Figure 18: Section of the chart of nuclides indicating isotopes that are especially suited for testing the unitarity of the CKM matrix. Red squares indicate ideal cases of Tz = 1 parent isotopes which decay to stable isotopes. Green and blue squares represent other good cases of Tz = 1 and Tz = 2 parent isotopes, respectively (Figure presented by G. Savard).*

Another example where the nucleus can serve as a laboratory to test fundamental models is the determination of $V_{ud}$, the up-down quark-mixing element of the CKM matrix. $V_{ud}$ can be determined from precise measurements of the ft-values for superallowed β transitions between analog $0^+$ states. These measurements have to be performed over a range of nuclides as the nuclear corrections factors necessary to extract $V_{ud}$ critically depend on several different nuclear quantities. Figure 18 shows the best candidates for these measurements. The determination of the ft-value requires measurements of the Q value, the half-life and the branching ratio to better than 0.1%. If the uncertainties of the evaluated data in ENSF are not of sufficient, dedicated experiments are required. It should be noted that the half-lives of the nuclides are the most precise half-life measurements in this mass region because of the dedicated ft-value experiments.

Finally, the nucleus can also be used to search for the permanent EDM. Placing limits on the EDM can establish strong constraints on models proposing physics beyond the Standard Model. In order to find the most promising nuclides for the observation of the EDM, their nuclear structure has to be known. For example, the presence of octupole vibrations can strongly amplify the sensitivity of a permanent electric dipole moment (EDM). Thus, the measurement and evaluation of octupole collectivity is an important component for nuclear data research related to the EDM search.

# 6. Dissemination

Previous sections have detailed the compiled and evaluated nuclear data that is needed for progress in a wide range of basic research in nuclear science. To utilize nuclear data in research, however, the data must be disseminated to the community. Effective dissemination can not only provide immediate positive impacts on research, it also provides a greater return on investment for resources to measure and evaluate the data.

Dissemination has evolved considerably over the last 50 years. ENSDF, for instance, started with paper publication in Nuclear Data Sheets, continued with remote terminal log in using Telnet, to finally reach interactive web applications. Most of the research community now uses web retrievals as a means to access nuclear data. As can be seen from Figure 19, there are more than 4 million retrievals from the NNDC web site (www.nndc.bnl.gov) per year and it is continuously growing. This site is the major dissemination center of nuclear data in the world, requiring a significant investment in hardware, software, and manpower to maintain its operation.

The most widely utilized gateway for providing ENSDF data is NNDC's interactive site NuDat (www.nndc.bnl.gov/nudat2). It provides displays of nearly 30 different quantities (lifetimes, separation energies, Q values, fission yields, and more) for more than 3300 nuclei. Clicking on a nucleus provides spin, parity, binding energy, lifetime, decay modes, and access to a list of levels, a level scheme, and decay radiation information. NuDat also provides search tools for γ-rays, energy levels, and decay radiation. In addition, the NNDC web site features many other means of disseminating nuclear data, including search and retrieval of nuclear structure and reactions files, specialized data sets (e.g., neutron resonances, safeguards decay standards). Of special note is the Nuclear Science

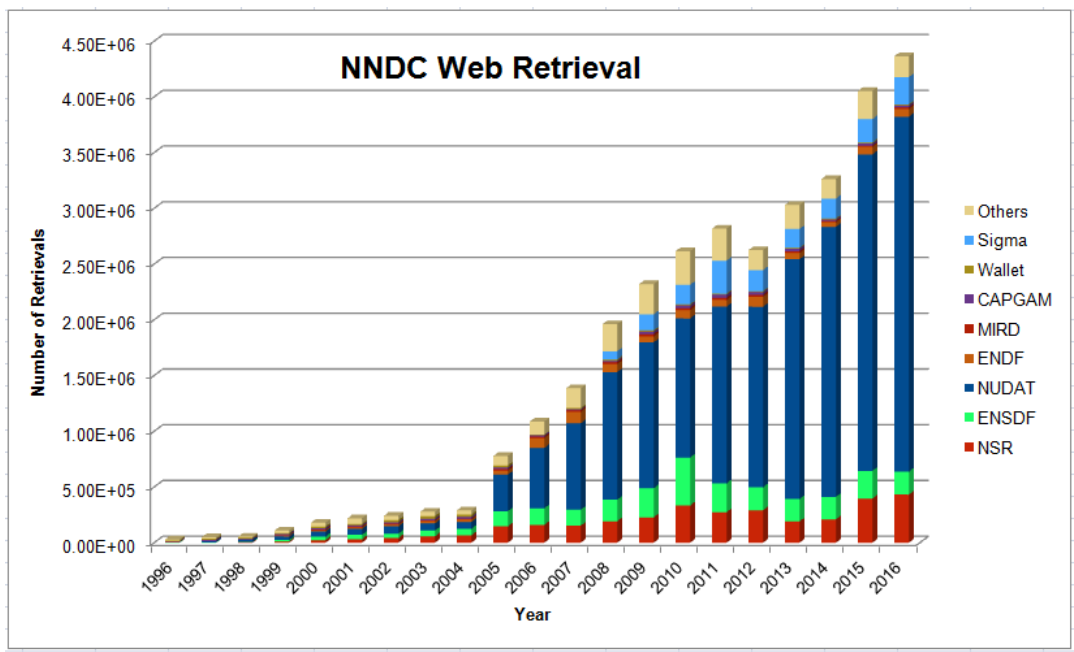

*Figure 19: Dissemination activities through the NNDC Web services.*

References (NSR), a continually-updated bibliographic database that covers nuclear physics research papers from over 80 journals.

Some new dissemination features have recently been implemented at the NNDC. For example, Figure 5 discussed above in the Nuclear Structure section, shows a plot of the ratio of the first $4^+$ to the first $2^+$ level energies for even-even nuclides. This ratio would be around 1 for spherical nuclides and 3.3 for deformed ones, showing clearly the increase of deformation away from magic numbers. Another new feature is the plot of nuclear level energy as a function of the angular momentum. This represents a new development, allowing a novel way to visualize level schemes. Figure 20 shows such a plot for $^{196}$Pb where the shape evolution as a function of angular momentum is apparent. The low-lying levels correspond to spherical shapes, while at higher excitation energies the levels indicate deformed and super-deformed shapes. This type of plot also reveals the existence of isomers and the levels built on top of them, as can be seen in Figure 21 for $^{178}$Hf where the 31-year $16^+$ isomer is clearly seen.

Similar new visualization tools should be developed for other nuclear properties including the possibility to plot correlations among them. Also – as mentioned earlier – there is the need to interface theoretical and evaluated databases to compare theory and experiment interactively.

Finally it is worth noting that the IAEA Nuclear Data Service (www-nds.iaea.org) serves as another major dissemination center for nuclear data. It offers an expansive collection of data bases and programs, some of which are partly based on the major USNP data bases. In addition, specialized USNDP web sites are hosted at ORNL (astrophysics at nucastrodata.org, nuclear masses at nuclearmasses.org, cosmology at bigbangonline.org) and TUNL (properties of light nuclei at www.tunl.duke.edu/nucldata).

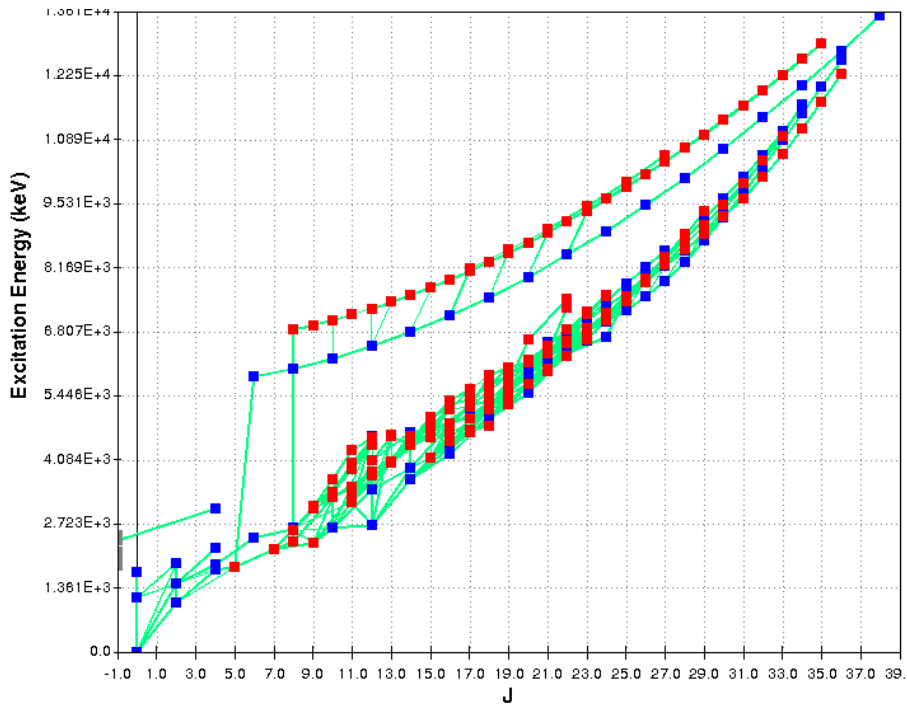

*Figure 20: Excited levels as a function of angular momentum joined by γ-ray branching ratios for $^{196}$Pb. The blue, red and grey squares represent positive, negative, and unknown parity, respectively.*

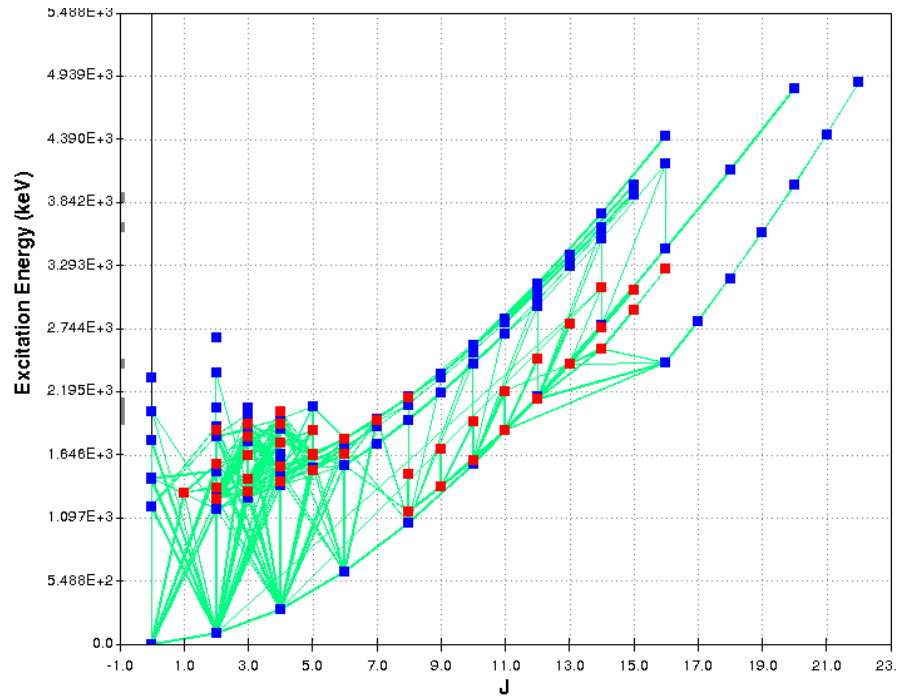

*Figure 21: Excited levels as a function of angular momentum joined by γ-ray branching ratios for $^{178}$Hf. The blue, red and grey squares represent positive, negative, and unknown parity, respectively.*

Common to all dissemination tools is that they should be easy to use, regularly updated, and they should continually evolve and expand in response to community needs. Development of new software tools and data collections, along with improvements of existing tools, would tremendously enhance the utilization of data disseminated by USNDP.

# Appendix 1: Historical Perspective

Compilation and evaluation of nuclear data have deep roots in nuclear physics research. The early compilations of known nuclides were published by Marie Curie et al. [1] and Giorgio Fea [2] in the 1930s. They were followed by the review article of M. Stanley Livingston and H.A. Bethe [3] where tabulated values of various reactions, decay properties and masses for stable and radioactive nuclei were presented. The compilation of nuclear properties continued with the publication of the Table of Isotopes series by G.T. Seaborg, J.M. Hollander, J.J. Livingwood, I. Perlman, and D. Strominger [4] in the journal *Reviews of Modern Physics*. In a letter to G.T. Seaborg in March 1941, the Assistant Editor of *Reviews of Modern Physics* stated *"I believe your suggestion of a revised list of radioactive isotopes for the April or July, 1942 issue of the REVIEW OF MODERN PHYSICS is a very good one. By that time the rate at which such radioactivities are discovered may be reduced very considerably and the table would itself become "stable""*. However, the amount of new data continued to grow substantially and the subsequent $6^{th}$ [5] and $7^{th}$ [6] editions of Table of Isotopes were completed at Lawrence Berkeley National Laboratory (LBNL) under the leadership of C.M. Lederer. The $8^{th}$ edition [7], which is the last in the Table of Isotopes series, was published in two volumes in 1996 by R. Firestone and co-workers. However, unlike previous publications, it was not an independent evaluation, but rather the data were mostly derived from the Evaluated Nuclear Structure Data File (ENSDF) database.

In 1945, in parallel to the Table of Isotopes effort, Katherine Way, who was John Wheeler's first graduate student, started collecting nuclear data as a part of her work for the Manhattan Project at Clinton Laboratory (later renamed Oak Ridge National Laboratory (ORNL)). In 1947, K. Way moved to the US National Bureau of Standards (later renamed US National Institute of Standards and Technology) in Washington, DC where she created the Nuclear Data Project in 1953. She coauthored a series of nuclear data reports that included basic properties of radioisotopes, such as half-lives, decay modes, energies and intensities of radiations, conversion coefficients, decay schemes and production methods. However, these reports did not provide any recommended values nor uncertainties. As the amount of data increased with time, those publications expanded considerably, but the data were still in the form of loose-leaf pages called Nuclear Data Sheets. In 1964, the Nuclear Data Project moved back to ORNL, once more under the leadership of K. Way, where the Nuclear Data Sheets were published in a booklet form. In February 1966, Nuclear Data Sheets became the section B of the journal *Nuclear Data*, and they were later published by Academic Press in a journal form, presently known as *Nuclear Data Sheets*. During that period of time, there was significant progress in both the development of comprehensive data formats and the computerization of the publication process for the Evaluated Nuclear Structure Data File (ENSDF) and Nuclear Science References (NSR) databases, thanks to the effort of E.B. Ewbank, M.J. Martin and co-workers at ORNL.

In late 1960s, both the Table of Isotopes Project at LBNL and the Nuclear Data Project at ORNL were having difficulty keeping up with the assessment of new data without additional resources. In the spring of 1969 an ad hoc panel on Nuclear Data Compilations was convened by the National Academy of Sciences to assess and evaluate the current situation and to make recommendations for reducing the backlog that has developed [8].

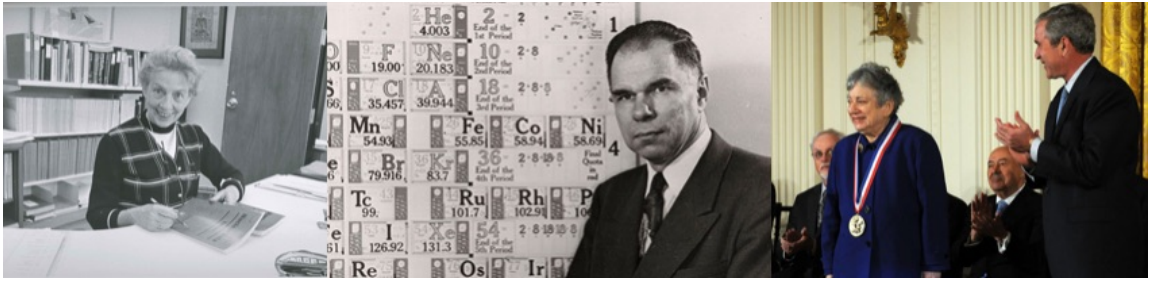

*Figure 22: From left to right: Katherine Way, Glenn Seaborg and Fay Ajzenberg-Selove*

It was recommended that the National Research Council of the academy conduct an intensive, two-year "crash" program with the main goal to get the nuclear data compilations up-to-date. This led to the recruitment of over 20 young, postdoctoral fellows who were selected to work on nuclear data evaluation in major nuclear structure laboratories in United States under leading nuclear experimentalists who served as sponsors. At the end of the program in 1974, several of these young scientists, including C. Baglin, T. Burrows, R.L. Bunting, E. (Gene) Henry and J. Tuli, remained as full time evaluators at several data centers in the United States. In mid 1970s under the guidance of Sol Pearlstein (NNDC), a national organization was created to coordinate the evaluation activities for mass chain evaluations among the national laboratories and universities in the United States. It included groups at BNL, ORNL, LBNL and Idaho National Engineering and Environmental Laboratory (INEEL), as well as scientists from Stanford University (W.E. Meyerhof, covering $A<5$), and University of Pennsylvania, (F. Ajzenberg-Selove, covering $5<A<20$ – later at TUNL). Another major development at that time was the relocation of the core of the program from ORNL to the National Nuclear Data Center (NNDC) at Brookhaven National Laboratory (BNL). Under the leadership of J. Tuli, NNDC assumed the responsibility for the maintenance of ENSDF and NSR databases, the corresponding analysis and production computer codes, and for the publication of the journal *Nuclear Data Sheets*.

In collaboration with the US nuclear data groups, compilation and evaluation work was also initiated in 1970 by M. Johns and his colleagues at McMaster University in Canada. Under the leadership of J. Kuehner, the McMaster nuclear physics group received in 1980 an independent grant from the Natural Sciences and Engineering Research Council of Canada for evaluation of nuclear structure data. The ENSDF evaluation work started that year with the participation of M. Johns and J. Ashbaugh, and under the guidance of B. Singh. In 1998, the McMaster data group became a member of the US Nuclear Data Program (USNDP) and received a direct grant from US DOE to support part of these activities. For many years B. Singh was at the core of the McMaster nuclear data effort with help from Emeritus Professor J. Cameron, who provided a volunteer service for about 10 years. Over the years, J.A. Kuehner, J.C. Waddington, D. Burke and A.A. Chen provided the much needed support to the nuclear data evaluation activities at McMaster University.

International collaborations came also as no surprise, since in addition to the activities in US and Canada, there were already independent evaluation effort in different parts of the world, including the work of B.S. Dzhelepov (and later with L. Peker, I.P. Selinov and others) in what was then the USSR, P. M. Endt and C. van der Leun at Utrecht University, and J. Blachot at CEA in Europe. Following on a recommendation by the International Nuclear Data Committee to the Nuclear Data Section (NDS) of the International Atomic Energy Agency (IAEA) in October 1973, a specialist IAEA meeting was held in the spring of 1974 in Vienna, Austria where it was proposed to expand the nuclear structure evaluation activity in order to include international groups [9]. A collaborative program was organized through the International Network of Nuclear Structure and Decay Data Evaluators (NSDD) under the auspices of the International Atomic Energy Agency (IAEA). This NSDD network began at a time when the workload was heavily reliant on USA input. A more equitable involvement of other national laboratories and universities from around the world was envisaged, and partially achieved. At different times prominent nuclear physicists such as F. Ajzenberg-Selove, R.G. Helmer, C.W. Reich, S. Raman (USA), J. Cameron (Canada), D. De Frenne, P.M. Endt, C. van der Leun, P.J. Twin and A.H. Wapstra (Europe) and many others, were involved in the resulting compilation and evaluation activities. Several countries have contributed over a long period of time, including Belgium, Canada, China, France, Japan, Kuwait, Russia, and the United States of America. Recently, new evaluation groups have emerged in other countries, such as Australia, India, Hungary and Romania.

The coordination of the NSDD network is the responsibility of the Nuclear Data Section of the IAEA. It involves organization of biennial meetings of the network and provides logistic support in the form of training workshops and coordinated projects aiming at improving the nuclear data evaluations and corresponding tools, such as analysis and checking codes, and training of new data evaluators.

[1] M. Curie, A. Debierne, A. S. Eve, H. Geiger, O. Hahn, S. C. Lind, St. Meyer, E. Rutherford, and E. Schweidler, Rev. Mod. Phys. **3**, 427 (1931).

[2] G. Fea, Nuovo Cimento **6**, 1 (1935).

[3] M. Stanley Livingston and H.A. Bethe, Rev. Mod. Phys. **9**, 245 (1937).

[4] J.J. Livingwood and G.T. Seaborg, Rev Mod Phys. **12**, 30 (1940); G.T. Seaborg, Rev Mod Phys. **16**, 1 (1944); G.T. Seaborg and I. Perlman, Rev Mod Phys. **20**, 585 (1948); J.M. Hollander, I. Perlman and G.T. Seaborg, Rev Mod Phys. **25**, 469 (1953); D. Strominger, J.M. Hollander and G.T. Seaborg, Rev Mod Phys. **30**, 585 (1958).

[5] Table of Isotopes, 6th Edition: C.M. Lederer, J. M. Hollander, and I. Perlman, John Wiley & Sons, Inc., 1968

[6]  Table of Isotopes, 7th Edition: C.M. Lederer, V.S. Shirley, E. Browne, J.M. Dairiki, R.E. Doebler, A.A. Shihab-Eldin, L.J. Jardine, J.K. Tuli and A.B. Buyrn, John Wiley & Sons, Inc., 1978

[7]  Table of Isotopes, 8th Edition: R.B. Firestone, V.S. Shirley, C.M. Baglin, J. Zipkin, S.Y. Frank Chu, John Wiley & Sons, Inc., 1996

[8]  Nuclear Data Compilations: The Lifeblood of the Nuclear Sciences and Their Applications, National Academy of Sciences, Washington DC, 1971

[9]  Summary Report on Specialists' meeting on Nuclear Data for Applications, Vienna, 29 April – 3 May, 1974; INDC(NDS)-060/W+spec.

# Appendix 2: Detailed Plan for Nuclear Astrophysics

Nuclear Astrophysics is focused on determining the cosmic origin of the elements and obtaining a better understanding of the nuclear physics driving the evolution and explosion of stars. This interdisciplinary field has extensive reaction and structure data needs combined with specialized data processing steps to provide critical input for simulations of cosmic systems. USNDP activities should be expanded to include efforts specifically targeted for nuclear astrophysics in order to maximize the scientific return on recent investments in this area.

## 1. Specific Needs and Capabilities for Nuclear Astrophysics

### 1.1 Overview

Nuclear astrophysics involves studies of some of the most fascinating systems in the Universe, from explosions like novae and X-ray bursts to the properties of neutron stars and their mergers, from the earliest stars in the universe to those millions of times more massive than the Sun, from Red Giants to White Dwarfs to Black Holes, from our Sun to the most distant stars. Such studies are a crucial component of the field of nuclear science, addressing one of the four overarching questions in nuclear science from the National Research Council's 2013 Assessment of Nuclear Physics and echoed in the 2015 Long Range Plan for Nuclear Science: "How did visible matter come into being and how does it evolve?" Determining the cosmic origin of the elements, and obtaining a better understanding of the nuclear physics that drives stellar evolution and stellar explosions, are the primary focus areas of this research. The importance of this work is confirmed in its role as a major motivation for the top priority construction project in U.S. nuclear physics, the Facility for Rare Isotope Beams (FRIB).

One of the most compelling aspects of this research is the inherent linking of the physics at the tiny distance scales of subatomic nuclei to stellar phenomena at scales $10^{24}$ times larger. Such a linkage requires extensive, diverse nuclear data sets that serve as the foundation for simulations of nucleosynthesis, stellar evolution, and stellar explosions. Predictions of these simulations have, in many cases, an extreme sensitivity to their input nuclear data.

These data needs cut across the traditional boundaries of nuclear structure and nuclear reactions. Furthermore, a specific set of processing steps is required wherein nuclear data is combined with nuclear theory and astrophysical theory to produce thermonuclear reaction rates. These rates are crucial, providing the foundation for all nucleosynthesis simulations. Without this specialized processing (*e.g.*, evaluation, extrapolation with theory, conversion to reaction rates), the latest experimental results and global nuclear theory calculations cannot be used in astrophysical simulations. The rapid flow of information between nuclear experimentalists and theorists, nuclear data evaluators, and astrophysics modelers is therefore essential for progress in this field.

## 1.2. Summary of Data Needs for Nuclear Astrophysics

Nuclear astrophysics has a wide range of nuclear data needs. We detail these needs below in the categories of nuclear reactions, nuclear structure and decay, processed data, software and dissemination, and other needs.

### 1.2.1. Nuclear Reaction Data Needs

Reaction cross sections play a key role in nuclear astrophysics, since it is their thermal average that generates the thermonuclear reaction rates upon which all simulations rely. The specific reaction data needs include:

- cross sections of (p,γ), (p,n), (p,α), (n,γ), (n,p), (n,α), and select ion+ion reactions on stable, proton-rich, and neutron-rich nuclei at energies ranging from ~ 100 keV to a few MeV in the center of mass
- theoretical cross sections via direct capture and statistical mechanisms, for the same reactions listed above, requiring optical model parameters, nuclear level densities, γ-ray strength functions, and predictions of the energies and spin-parities of bound levels
- cross sections of photonuclear reactions on stable and proton-rich nuclei above Fe
- cross sections of neutrino interactions on nuclei in the Fe-group, on those near $^{64}$Ge, and those in the sd-shell

### 1.2.2. Nuclear Structure and Decay Data Needs

Level properties are needed to estimate the cross sections of reactions where no direct measurement is available. Beta- and electron capture decays are critical because in many astrophysical environments, the competition between capture reactions and decays towards stability determines the reaction pathways and therefore the nuclei synthesized in these events. Nuclear masses are essential for the determination of energy release in thermonuclear reactions, determination of the ratios of forward and reverse rates strongly affecting for example simulations of r-process nucleosynthesis occurring in supernovae that follow paths of constant neutron separation energy, global models of the structure of nuclei, and statistical model calculations of reaction cross sections. Fission barriers are needed to understand neutron-induced fission occurring in supernovae and neutron star mergers. A detailed list of the structure and decay data needs includes:

- properties of low-lying single particle levels including excitation energies, spectroscopic factors, Asymptotic Normalization Coefficients (ANCs), spin parities, partial gamma and particle widths, total widths, and resonance strengths
- nuclear masses for the entire nuclear chart
- weak interaction strengths including β/positron decay, electron capture, β-delayed proton decay, β-delayed single- and multiple-neutron decay, branching ratios
- fission barriers and of the mass distribution of fission products

### 1.2.3. Specialized Nuclear Data Sets

While some of the required nuclear data are available as a part of the current US data program, many of the most critical ones are not. Thermonuclear reaction rates are the foundation of all nuclear astrophysical simulations. Many are processed from the thermal averaging of cross sections with extrapolations to astrophysical energies, while others are based on analytical calculations using the properties of known and predicted levels near particle thresholds or statistical models. Often these approaches have to be combined. The steps needed to generate reaction rates are specific for each reaction type. Since there are typically hundreds to thousands of thermonuclear reactions relevant for studies of any given astrophysical environment, the collection of rates into libraries is essential for progress in the field. The nuclear astrophysical community has advocated the uniform adoption of the JINA REACLIB library as the standard, with new smaller libraries created by research groups being folded into JINA REACLIB in order to be distributed to the community. Strong support for continuing to grow this library, as well as the associated software tools used to manage, modify, and share the rates, is essential for progress in the field.

There are other needs as well. Astrophysical data such as measured or observed abundances are important as initial values for simulations and as the reference points to which simulation predictions are compared. Partition functions and stellar enhancement factors (SEFs) are used to correct laboratory-measured cross sections (usually performed on nuclei in their ground state) to include reactions on the fraction of nuclei in a star that are thermally excited into low-lying excited states. The nuclear equation of state, essential for modeling astrophysical environments with the highest temperatures and densities, relies on the compressibility of nuclear matter as a crucial constraint. Weak interaction rates are important in almost all nucleosythesis processes. The specific needs are:

- thermonuclear reaction rates for $(p,\gamma)$, $(p,n)$, $(p,\alpha)$, $(n,\gamma)$, $(n,p)$, $(n,\alpha)$, and select ion+ion reactions on stable, proton-rich, and neutron-rich nuclei over a variety of mass ranges at temperatures from 1e07 - 1e10 K
- partition functions and stellar enhancement factors for these reactions
- nuclear equation of state (EOS) and nuclear compressibility
- astrophysical data including abundances of nuclei in a variety of cosmic systems
- electron capture and beta decay rates as functions of electron density and temperature

### 1.2.4. Software and Dissemination Needs

For the community to fully utilize the datasets mentioned above, the availability of appropriate processing codes, software tools, and dissemination services is essential. A USNDP effort at ORNL produced the Computational Infrastructure for Nuclear Astrophysics, a unique online system successfully that contains many of the processing tools listed below. This system is used by researchers in 160 institutions in 35 countries. Continuous maintenance and development of this system will be important. A list of tools needed by the community include:

- online tools to generate, modify, manage, storage, merge, customize, visualize, and share thermonuclear reaction rates

- online codes to benchmark reaction rate libraries
- online tools for customized plots of cross sections, rates, benchmark calculations, and related data sets
- online and downloadable codes for structure and reaction calculations
- search engines spanning multiple databases
- online and downloadable codes to generate data set uncertainties

### 1.3. Prioritization

Since the needs listed above are extensive, it is important to indicate the highest priority items needed for astrophysical studies. The priority items listed below reflect current scientific emphasis, needs related to recent investments in stable and unstable beam facilities, and areas where additional effort is most urgently needed to take full advantage of experimental data:

- ongoing evaluation of thermonuclear reaction rates using recent and new experimental and theoretical data.
- reliable temperature and density dependent electron capture and beta-decay rates across the chart of nuclides using a range of experimental and theoretical data

Also of critical importance is the continuation of the ongoing atomic mass evaluation efforts, and evaluation of experimental beta decay and electron capture rates, including branchings of particle emission.

### 1.4. Current Situation

The data needs mentioned above, some of which are specific to nuclear astrophysics, have to a large extent fallen through the gaps of traditional efforts in the nuclear data community. As a result, progress in nuclear astrophysics has been hampered by extensive, unmet nuclear data needs. In recent years, the nuclear astrophysics research community responded by launching a number of small scale projects involving reaction assessments (*e.g.*, the NACRE and KADONIS efforts in Europe), thermonuclear rate libraries (*e.g.*, the JINA REACLIB rate library), and a Monte Carlo approach for rate uncertainty estimation (*e.g.*, STARLIB at Univ. North Carolina). The USNDP had also previously supported the Computational Infrastructure for Nuclear Astrophysics at ORNL. While these efforts represent significant progress in the field, they lacked continuity and longevity, as well as sufficient manpower in evaluations to keep pace with measurements. As a result, many new results in low-energy nuclear physics cannot yet be used in astrophysical simulations. This situation will worsen as new facilities (e.g., FRIB, the CASPAR underground facility, St. George at Notre Dame, upgraded LENA at UNC) and new experimental devices (e.g., SECAR) come online. The scientific potential of these large investments in experimental nuclear astrophysics will not be fully realized without a parallel increase in efforts targeted at providing the required nuclear data.

## 1.5. Path Forward

There is a strong need for a dedicated effort by the USNDP in nuclear astrophysics data which can give the required continuity, longevity, and manpower boost to enable progress in this field. Such an effort can maximize the science impact of the latest investments in major facilities and equipment in the nuclear science program. Such an effort

- should be closely coordinated with the broader nuclear astrophysics community to ensure the specific needs of this field are met
- should build on existing efforts
- should be spread across different institutions
- should focus on assessments of data most critical for astrophysics simulations
- should include efforts to develop and spread the special expertise required for this work
- must include development and maintenance of software and dissemination tools
- should include robust storage solutions
- should be flexible to evolve as data needs change in the field
- should emphasize consistency across the field

We therefore recommend that

[1] the USNDP mission statement be expanded to specifically mention nuclear astrophysics

[2] USNDP activities should be expanded to include meeting the astrophysics data needs detailed above, with efforts specifically targeted for nuclear astrophysics

As a first implementation, we recommend the following steps:

[A] Enhance evaluation of astrophysical reactions by incorporating this into the work plan at the MSU Data effort and other potential evaluation efforts

[B] Enhanced software tool development and maintenance by re-instating this activity in the ORNL Nuclear Data Project

[C] Initiate the development of next generation reaction rate libraries by supporting work on STARLIB at the UNC data Center that would work closely with the JINA REACLIB effort at MSU

This will be the start of a coordinated U.S. effort to produce and disseminate critically important, high-quality evaluations of nuclear data for nuclear astrophysics. Such an effort will have a significant positive impact on progress in nuclear astrophysics research.

# Appendix 3: Survey

**Survey of Nuclear Data: Databases and Applications**

---

How often do you use the databases?

_____ NSR  _____ ENSDF  _____ XUNDL  _____ EXFOR  _____ ENDF

(1) Every day  (2) Once a week  (3) Once a month  (4) Rarely  (5) Never heard of it

---

How would you rate the quality of?

_____ NSR  _____ ENSDF  _____ XUNDL  _____ EXFOR  _____ ENDF

(1) Excellent  (2) Good  (3) Average  (4) Poor  (5) No opinion

---

Does the currency of the database meet your research needs for?

_____ NSR  _____ ENSDF  _____ XUNDL  _____ EXFOR  _____ ENDF

(1) Yes, all the time  (2) Yes, mostly  (3) No, needs more frequent updating  (4) No opinion

---

How would you rate the user-friendliness of the NNDC website?

___ Excellent  ___ Good  ___ Needs Work  ___ Can't find anything

Are current web-searching tools sufficient for your research? ___ Yes  ___ No

If no, what you like to be able to search for? ______________________________

---

Do you download pdf versions of the Nuclear Data Sheets Journal?  ___ Yes  ___ No

If yes, how often?  ___ Once a year  ___ Once a month  ___ Once a week

Do you use the printed copy of the Nuclear Wallet Cards?  ___ Yes  ___ No

---

How do you search for publications? ___ NSR  ___ WebofScience  ___ inSPIRE  ___ Other

How do you search for Nuclear Structure Data (check all that apply)?

___ NuDat  ___ ENDSF Web Retrival  ___ Chart of Nuclides  ___ XUNDL  ___ Other

How do you search for Nuclear Reaction Data (check all that apply)?

___ Sigma  ___ ENDF Web Retrival  ___ EXFOR  ___ Other

---

General Comments and Suggestions

[ ]

## How often do you use the databases?

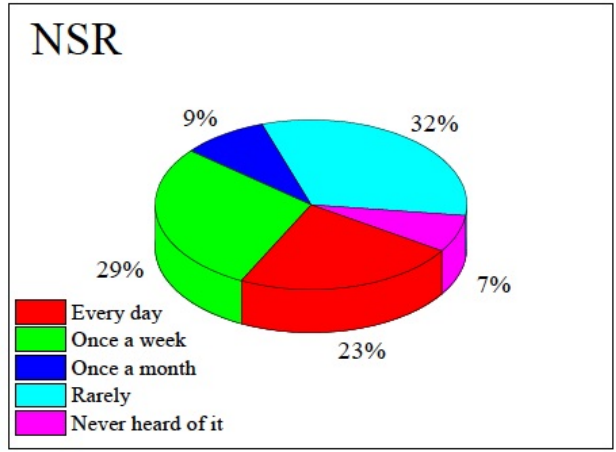

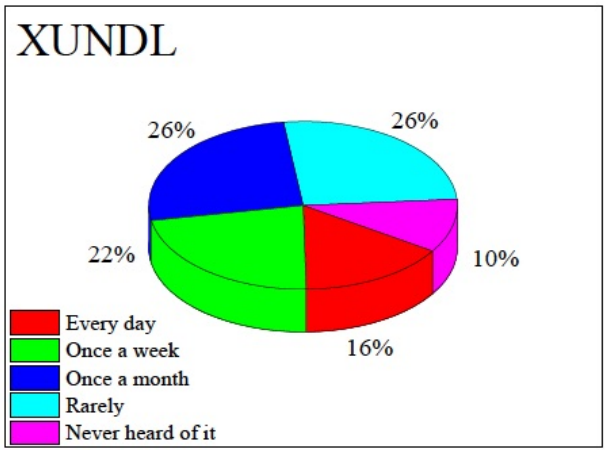

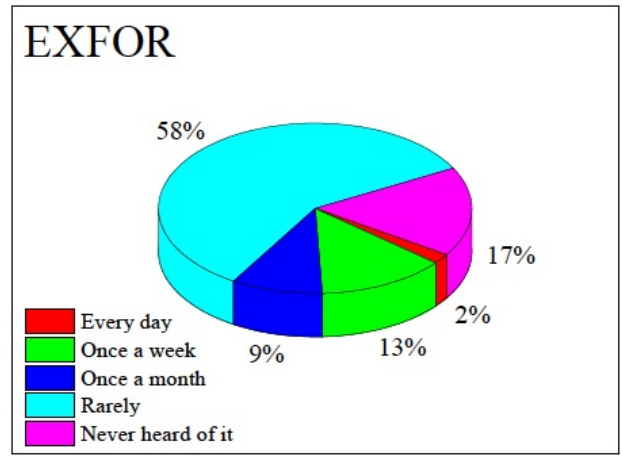

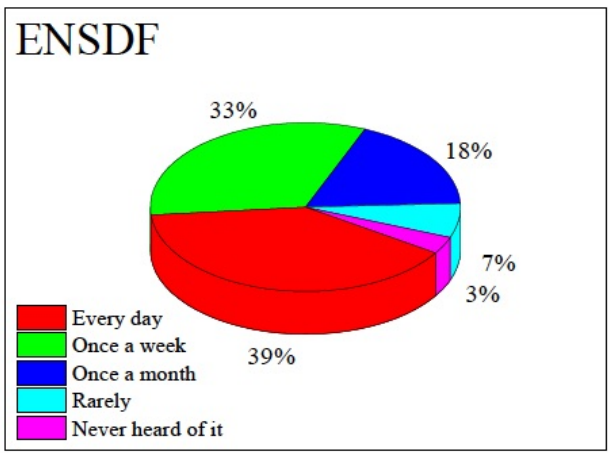

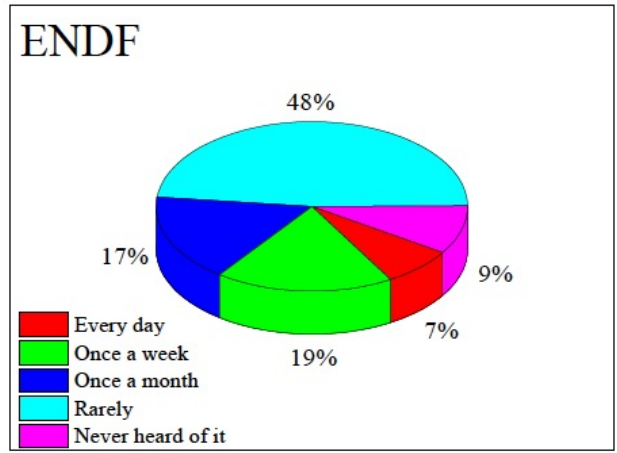

**How would you rate the quality of the databases?**

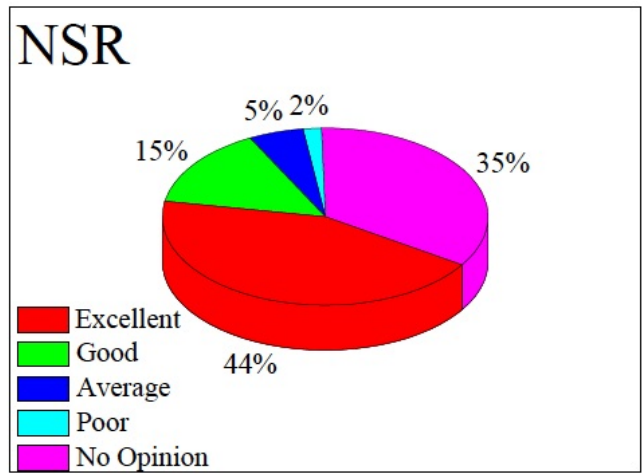

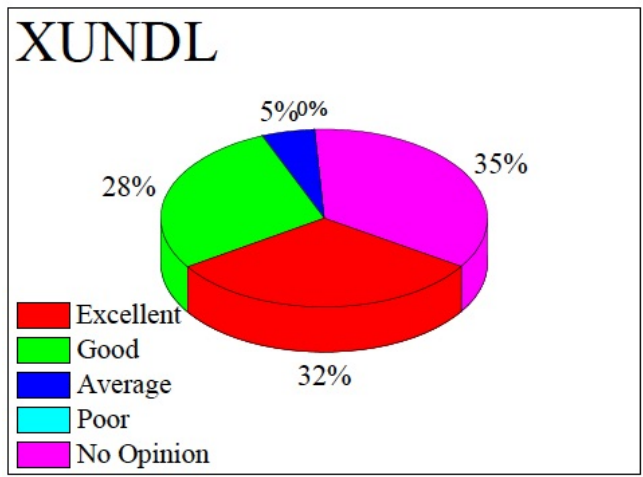

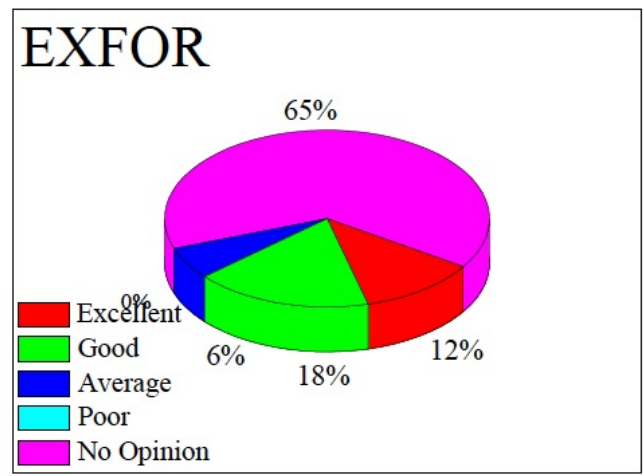

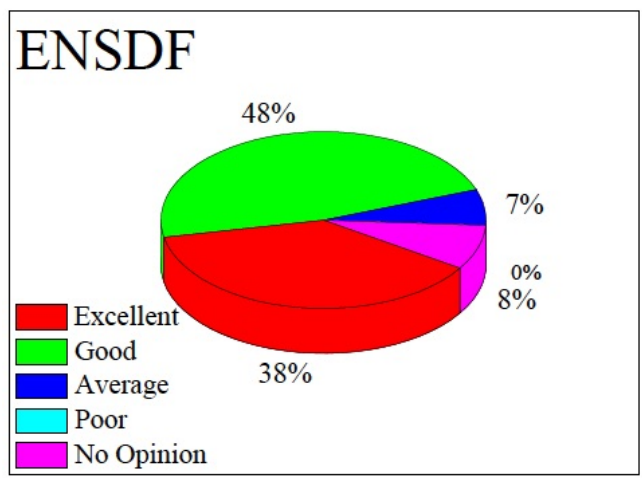

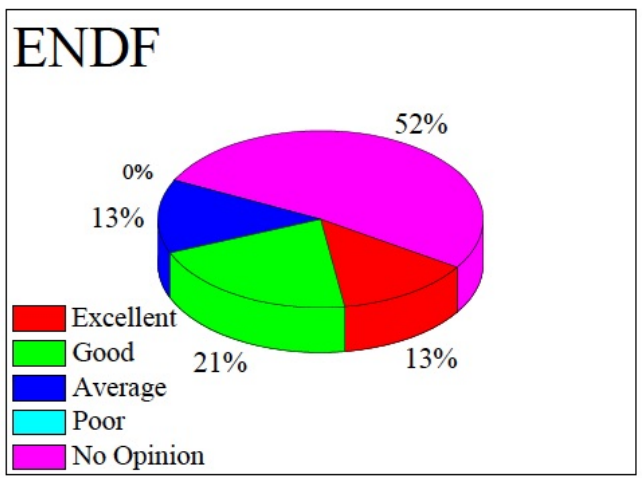

**Does the currency of the databases meet you need?**

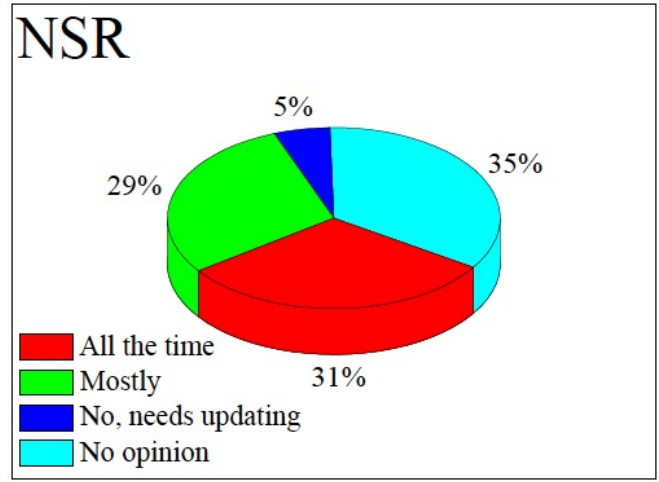

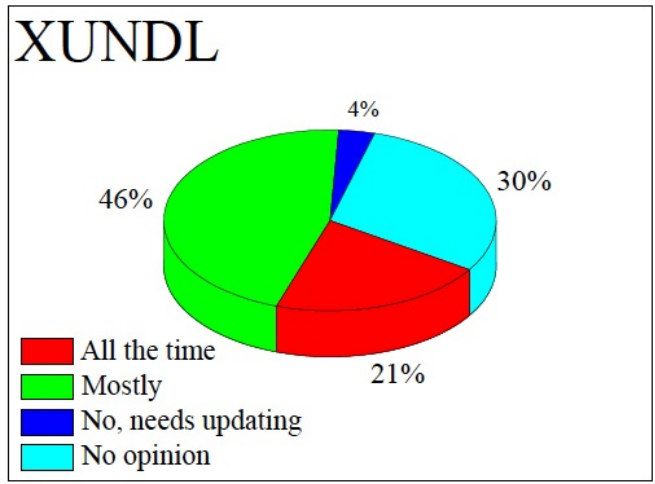

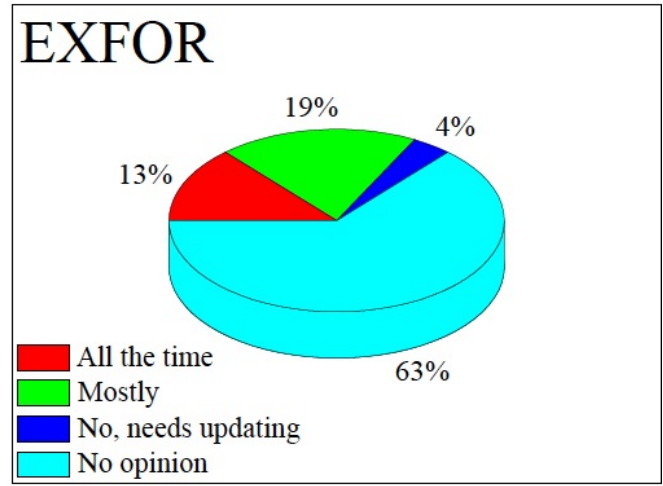

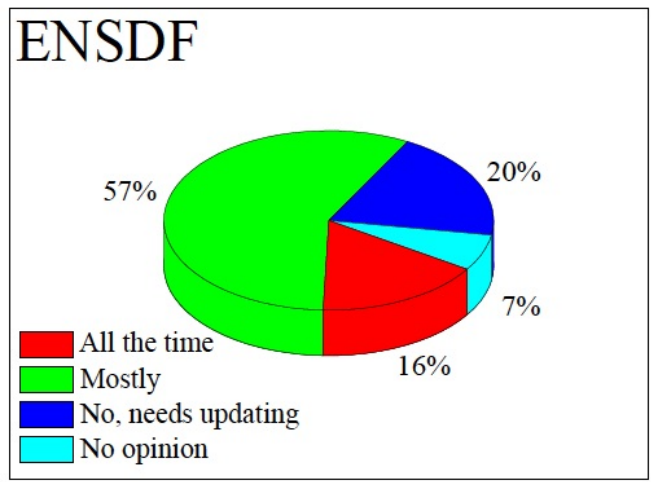

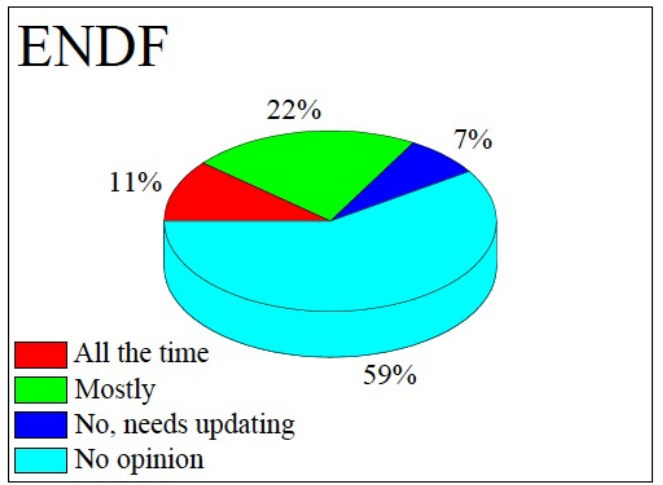

**How would you rate the user-friendliness of the NNDC Website?**

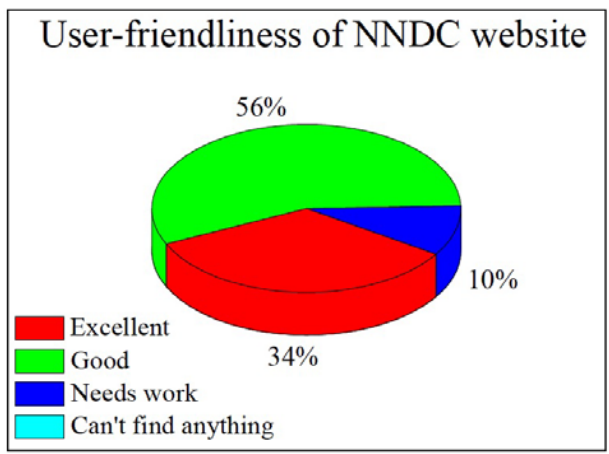
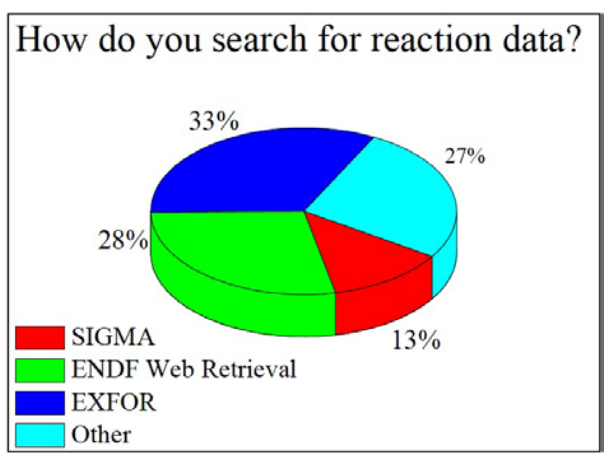
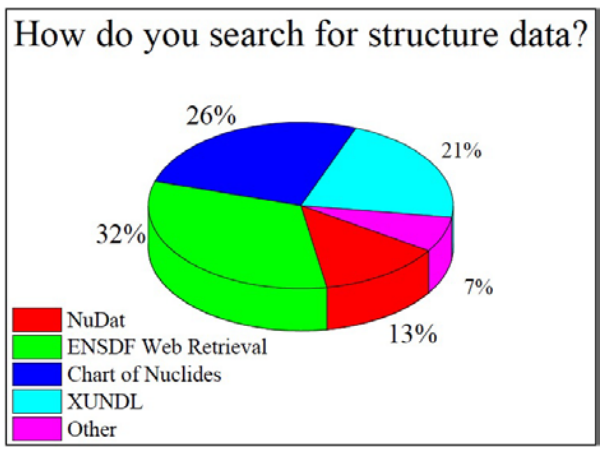
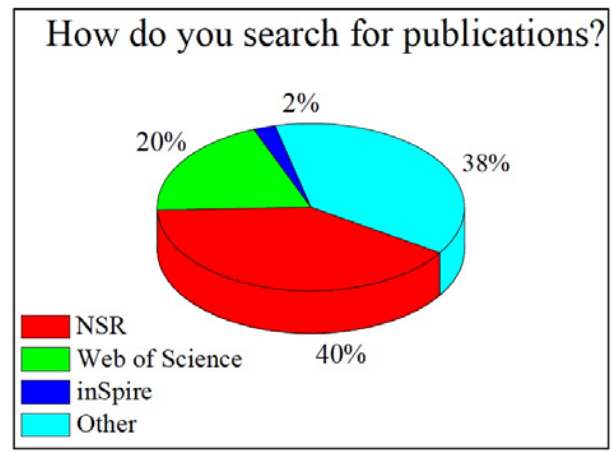